\documentclass[breakmath, preprint]{seismica}

\usepackage{dblfloatfix}     % fixes some twocolumn float placement issues

\title{Exploration of Machine Learning Methods to Seismic Event Discrimination in the Pacific Northwest}
\shorttitle{Seismic event discrimination in the PNW} % used for header, not mandatory

\author[1]{Akash Kharita
\orcid{0000-0003-0612-7734}
\thanks{Corresponding author: ak287@uw.edu}
}
\author[1]{Marine Denolle
\orcid{0000-0002-1610-2250}
}

\author[1,2]{Alexander R. Hutko
    \orcid{0000-0003-3561-2747}}
    
\author[1,2]{J. Renate Hartog
    \orcid{0000-0002-4116-7806}}
    
\author[1,2]{Stephen D. Malone
    \orcid{0000-0002-5143-1850}}

\affil[1]{Earth and Space Sciences, University of Washington, Seattle, USA}
\affil[2]{Pacific Northwest Seismic Network, University of Washington, Seattle, USA}

\credit{Conceptualization}{AK,MD}
\credit{Methodology}{AK, MD}
\credit{Software}{AK}
\credit{Validation}{AK,AH,SM}
\credit{Formal Analysis}{AK,MD}
\credit{Investigation}{AK,MD,AH}
\credit{Resources}{RH}
\credit{Writing - Original draft}{AK}
\credit{Writing - Review \& Editing}{AK,MD,AH,SM}
\credit{Visualization}{AK}
\credit{Supervision}{AK,MD,RH}
\credit{Project administration}{MD,RH}
\credit{Funding acquisition}{MD,RH}

\begin{document}

%% Your article can include up to 3 abstracts. The first is the English language abstract.
%% For other languages in the second and third optional abstracts, you might have to define 
%%      additional font(s) in preamble above
%% You can also include a non-technical summary in addition to the abstract(s)
\maketitle
\begin{summary}{Abstract} % 174 words < 200

Accurately separating tectonic, anthropogenic, and geomorphologic seismic sources is essential for Pacific Northwest (PNW) monitoring but remains difficult as networks densify and signals overlap. Prior work largely treats binary discrimination and seldom compares classic ML (feature-engineered) and deep learning (end-to-end) approaches under a common, multi-class setting with operational constraints. We evaluate methods and features for four-way source discrimination—earthquakes, explosions, surface events, and noise—and identify models that are both accurate and deployable. Using $\sim$200k three-component waveforms from $>$70k events in an AI-curated PNW dataset, we test random-forest classifiers on TSFEL, physics-informed, and scattering features, and CNNs that ingest time series (1D) or spectrograms (2D); we benchmark on a balanced common test set, a 10k-event network dataset, and out-of-domain data (global surface events; near-field blasts). CNNs taking spectrograms lead with accuracy performance over 92\% for within-domain (as a short-and-fat CNN SeismicCNN 2D) and out-of-domain (as a long and skinny CNN QuakeXNet 2D), versus 89\% for the best random forest; performance remains strong at low SNR and longer distances, and generalizes to independent network and global datasets. QuakeXNet-2D is lightweight ($\approx$70k parameters; $\sim$1.2 MB), implemented into {\tt seisbench}, scans a full day of 100 Hz, three-component data in \~9 s on commodity hardware, with released checkpoints. These results show spectrogram-based CNNs provide state-of-the-art accuracy, efficiency, and robustness for real-time PNW operations and transferable surface-event monitoring.

\end{summary}

\section{Introduction}

\begin{figure*}
\includegraphics[width=\textwidth]{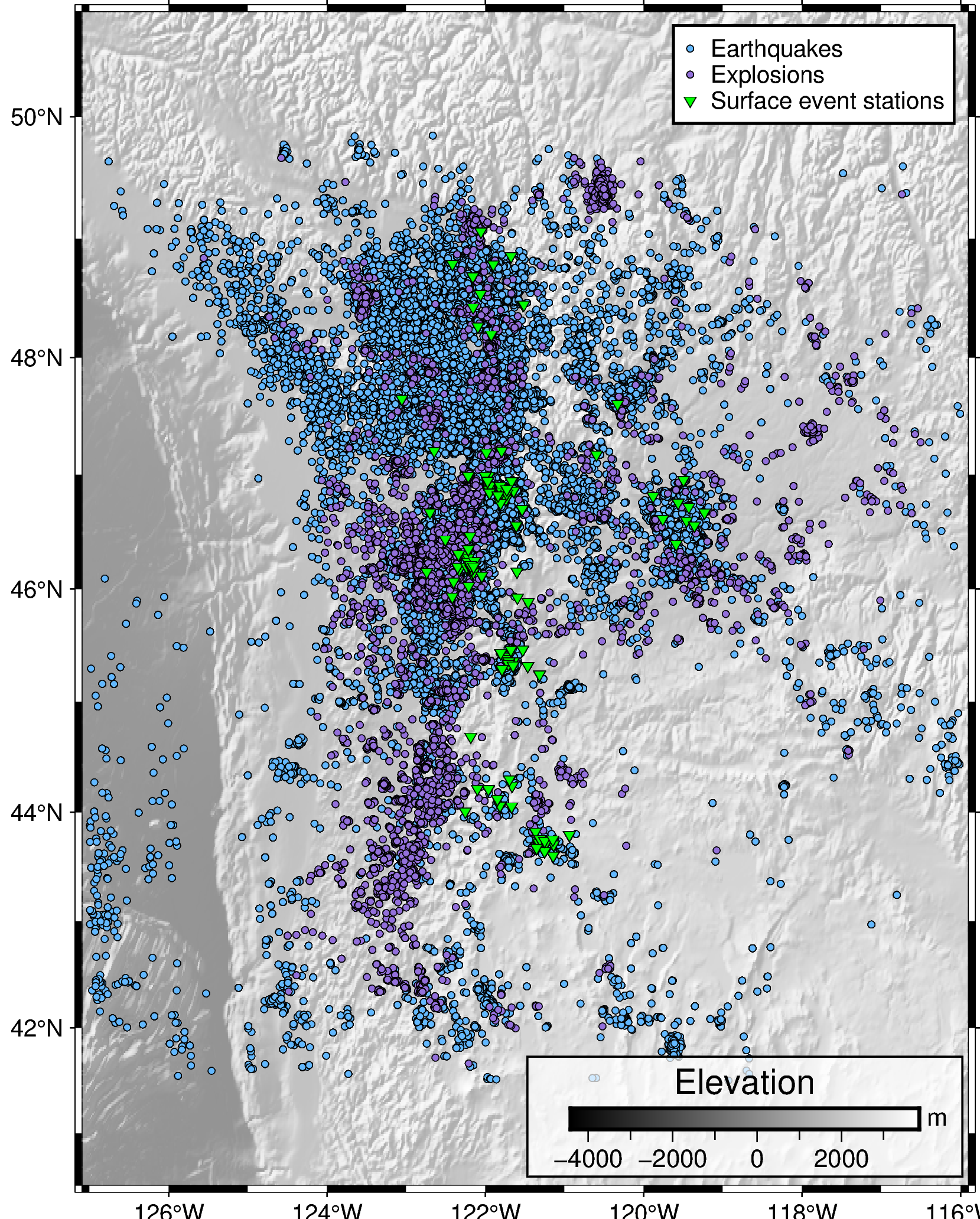}
\caption{\textbf{Map of seismic events in the curated catalog by \citet{ni2023curated}}. Earthquakes (blue circles) and explosions (purple) are located by the PNSN. Surface events are only marked at the seismic stations where they are recorded (green triangles).}
\label{fig:figure1}
\end{figure*}

The Pacific Northwest (PNW) region of the United States, situated at the dynamic boundary between the North American continental plate and the Juan de Fuca oceanic plate, presents unique challenges and opportunities in seismic monitoring in a multi-geohazard-prone landscape with a subduction-zone plate boundary. The PNW experiences diverse seismic sources (Fig.~\ref{fig:figure1}), including large megathrust earthquakes \citep[e.g.,][]{witter2003great}, intraslab \citep[e.g.,][]{ichinose2004rupture} and crustal earthquakes \citep[e.g.,][]{gomberg2021productivity}, slow repeating earthquakes \citep[e.g.,][]{bartlow2020long, rogers2003episodic, wech2014slip}, tectonic tremors \citep[e.g.,][]{wech2010earthquake}, and low-frequency earthquakes \citep[e.g.,][]{royer2014comparative}. Beyond earthquakes, over twenty active and glaciated volcanoes, as well as extensive mountain ranges, experience frequent landslides and debris flows \citep[e.g.,][]{luna2022seasonal}. Anthropogenic activities, such as quarry blasts (REF), generate ground motion intensities comparable to those of small-magnitude earthquakes, further complicating the source of this seismicity (REF)\citep{kramer2024recent}. Such a variety of seismic sources necessitates robust classification methods to accurately label and catalog these events . 

% seismic monitoring in the PNW
The PNW Seismic Network (PNSN) \citep{hellweg2020regional}, a key component of the Advanced National Seismic System (ANSS), has been operating since 1969 and currently manages over 600 seismic stations in the states of Washington and Oregon, providing essential data for seismic event analysis. Current event detection relies on traditional techniques such as the Short-Time Average to Long-Time Average (STA/LTA) ratio algorithm \citep[e.g.,][]{allen1982automatic}. While effective for basic event detection, this approach has limited accuracy when discriminating between visually similar waveforms from different event types, such as earthquakes, controlled explosions, and glacier-related surface events. These limitations have become more pressing with the expansion of seismic networks and the increasing volume of data \citep[e.g.,][]{carniel2021machine, kong2019machine}, particularly in regions like the PNW, where the simultaneous occurrence of multiple seismic sources adds to the complexity of waveform interpretation. Traditional discrimination techniques, such as analyzing the spectral ratios of P and S waves or differences in local and coda magnitudes, which were developed for binary classification between earthquakes and explosions \citep[e.g.,][]{koper2016magnitude,Koper24}, also face limitations. For example, these parameters are often unavailable during preliminary analyses, which prevents accurate classification in near real-time. While calculating duration (coda) magnitude and local magnitude alongside their spectral ratios could provide valuable insights into potential misclassifications \citep[e.g.,][]{koper2016magnitude,Koper24}, such comprehensive analyses are rarely performed in real-time workflows.

% where we expect the confusion between the events? What is the magnitude and depth and epicentral distance range where these events overlap? What can be confusing between surface event waveforms and volcano related seismicity?

\begin{figure*}
\includegraphics[width=\textwidth]{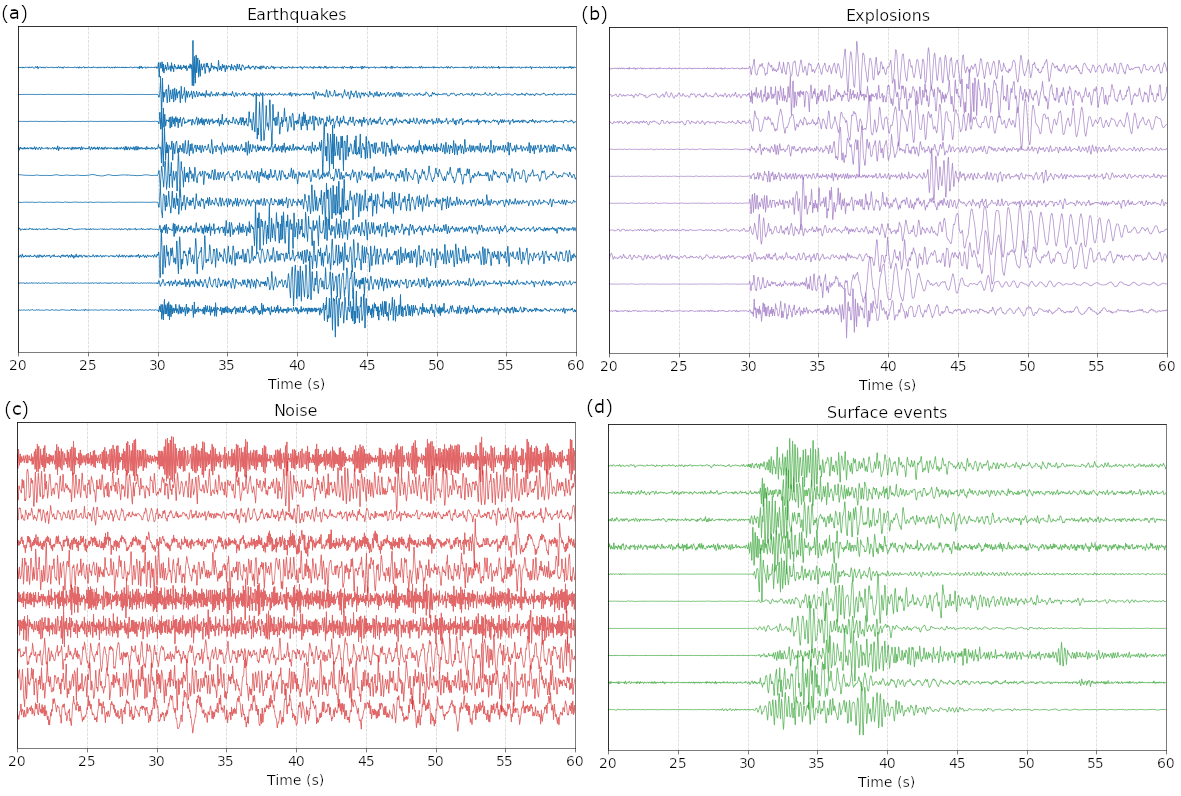}
\caption{{\bf Random examples of waveforms in each class.}}
\label{fig:figure2}
\end{figure*}

Small magnitude earthquakes (M$_L$ < 3) have been difficult to distinguish from mining and single-shot explosions recorded at local distances (< 50–150 km) due to similarities in their seismic waveforms and spectral characteristics ( Fig.~\ref{fig:figure2}) and (Fig.~\ref{fig:figure3}) particularly at frequencies below 10 Hz \citep[e.g.,][]{wang2020seismic, koper2021discrimination}. Similarly, seismic signals generated by mass movements on volcanoes, such as landslides, debris flows, and lahars, closely resemble those associated with low-frequency volcanic seismicity and shallow volcano-tectonic earthquakes. These events are characterized by emergent waveforms, indistinguishable P and S phases, and dominant frequencies below 5 Hz \citep[e.g.,][]{wassermann2012volcano, allstadt2014seismic, allstadt2018seismic}. Furthermore, seismic signals from rockfalls with a significant free-fall component can exhibit similarities to small earthquakes when recorded locally \citep{hibert2011slope, hibert2014automated}. In volcanic regions near populated areas, volcano-seismic signals, including those related to low-frequency volcanic events or shallow volcano-tectonic earthquakes, can also be misinterpreted as anthropogenic noise \citep{wassermann2012volcano}.

% one paragraph on how many studies have tackled the binary classification, especially of earthquakes and explosions, or earthquake and slope failures, or earthquakes and icequakes, but that complex regions such as volcanic regions have necessitated the multi-classification problem (cite multi class projects). Discuss how binary classification is easier (higher reported F1 scores) relative to multiclass classification (in average lower F1 scores?).

Many studies have focused on the binary classification of seismic events, particularly distinguishing between earthquakes and explosions \citep[e.g.,][]{koper2021discrimination, kong2022combining}, or other sources such as slope failures \citep[e.g.,][]{wenner2020near,chmiel2021machine, hibert2017automatic} or icequakes \citep[e.g.,][]{Pirot23,kharita2024discrimination}. Binary classification is generally easier because it involves separating only two classes with more distinct waveform features and requires a solution for a single decision boundary. As a result, performance in binary problems is typically very high, with many studies reporting classification accuracies exceeding 95\% \citep[e.g.,][]{wang2020seismic, koper2021discrimination}.  

By contrast, in complex environments such as volcanic regions, where seismic signals from landslides, pyroclastic flows, and low-frequency volcanic events share similar waveforms, multi-class classification is necessary and considerably more difficult \citep[e.g.,][]{wassermann2012volcano, allstadt2014seismic, allstadt2018seismic}. Discriminating among multiple event types is more challenging because the boundaries between classes are less distinct, and performance is typically lower, often in the range of 75–90\% \citep[e.g.,][]{hibert2014automated, hibert2017automatic, hibert2019exploration}.

Classic Machine Learning (CML) and Deep Learning (DL) have shown promise in seismic event classification by enabling automated feature extraction and effective discrimination between event types, even in noisy environments. CML techniques, which use engineered features as input (see Fig.~\ref{fig:figure4}), may offer interpretability and have shown success in distinguishing between event classes such as earthquakes, explosions, and glacier seismicity \citep[e.g.,][]{koper2016magnitude, zeiler2009developing, kharita2024discrimination, hibert2017automatic, pirot2024enhanced, domel2023event, wang2023using}. However, these approaches often require extensive and computationally costly feature engineering, which may limit adaptability to new event types. DL methods, on the other hand, automatically extract features from raw data through neural network optimization (see Fig.~\ref{fig:figure4}), resulting in good classification performance for more nuanced differences in seismic signals \citep[e.g.,][]{mousavi2022deep, bergen2019machine}. Studies have typically chosen either CML or DL approaches and used varied data sets, which limits our ability to draw a general understanding of feature extraction and machine learning classification on multi-class discrimination.

%Review of interpretability of deep learning models, necessary on generalizability: Kong, Linville, Ross? grad-CAM, LRP, activation layer 
%Volcano seismology mostly uses random forest for feature importance. 
The interpretability of machine learning models is crucial for understanding and ensuring the generalizability of their predictions. In CML, this is typically achieved by exploring feature importance, which often reveals that certain seismic wavefield features are effective in discriminating between events of different classes. Knowing which features are most important, scientists have the opportunity to investigate the varied physical processes that generate the differences in features (i.e., the seismic wavefield signatures). The interpretability of deep learning models is less straightforward, mostly because isolating parts of the feature space that most contribute to the classification is buried in neural networks. There exist methods today to estimate feature importance in seismological applications. For example, \citet{kong2021deep} and  \citet{kong2022combining} used a method called Grad-CAM to trace back regions (time and feature space) of the seismic waveforms that most contribute to classification, \citet{linville2019deep} used attention mechanisms to identify key temporal patterns in seismic signals, \citet{clements2024grapes} directly visualized a given feature value after activation to isolate the wave types that contribute to improving shaking intensity forecast for early warning. A comprehensive comparison between CML-based and DL-based methods is still lacking.

 % For example, \citet{kong2021deep} and  \citet{kong2022combining} used a method called Grad-CAM to trace back regions (time and feature space) of the seismic waveforms that most contribute to classification, helping to build confidence in the model’s decisions if these regions point to known phases of the wavefield, which were intuitive in the context of discrimination between earthquake and explosions. \citet{linville2019deep} used attention mechanisms to identify key temporal patterns in seismic signals, enhancing model interpretability for the classification of volcanic seismic events, such as eruptions and volcano-tectonic earthquakes.  A comprehensive comparison between CML-based and DL-based methods is still lacking. 

Accurate multi-class classification in the PNW is not only a technical challenge but also a scientific and operational priority. The region’s overlapping seismic sources generate visually similar signals, making them difficult to distinguish in real time. Improving classification reduces analyst workload, enhances the reliability of catalogs, and supports downstream applications such as hazard assessment, early warning, and tomography. Beyond operations, clean catalogs enable new scientific insights—for instance, clarifying the physical differences between explosions, earthquakes, and mass movements, or quantifying the seasonal and climatic controls on surface processes. In particular, building the first comprehensive catalog of surface events in the PNW would provide a baseline for future research on landslide frequency–magnitude statistics and volcano–geomorphic interactions.

The unique contribution of this work is the comprehensive evaluation of waveform feature space and model architectures (CML and DL) for the multi-class classification of seismic events. In Section 2, we describe the data set compiled for this study, which includes the curated data by  \cite[e.g.,][]{ni2023curated} with its diverse tectonic, anthropogenic, and geomorphological seismic sources, the exotic event catalog compiled by \citet{bahavar2019exotic}, and additional data we prepared for model development and testing. Section 3 describes the methodologies, both CML and DL, the workflows downstream to deploy these models, performance evaluation on unseen data from the PNSN, and generalization assessment with within and out-of-domain data. Section 4 presents the model's performance. Section 5 explores into feature importance estimated from these two machine learning approaches. Additional points of discussion are in Section 6, including an analysis of the misclassified events, context for model performance with respect to other published work, a use case that deploys the best-performing classifier on continuous data to detect events, and a suite of recommendations for using these models.

% generalization of the model with a focus on surface events by testing it on global data sets of mass movements. In section 6.4, we delve into the operationalization of the classifier at a regional seismic network, PNSN, by deploying and testing the model performance on realistic, imbalanced
% considerations in the datasets. Finally, we test and discuss in section 6.5, computational considerations for running the classifier at scale on continuous data by embedding a Python ecosystem for earthquake catalog building {\tt seisbench}  \citep{Woollam22}, and the scalability of deploying the detection at scale. 

\section{Data}
This study utilizes multiple datasets to train and test the models. The first is a subset of the comprehensive dataset curated by \citep{ni2023curated}, which spans 21 years, from 2002 to 2022, and additional waveforms we collected. The curated dataset comprises approximately 200,000 seismic waveforms and associated metadata, corresponding to approximately 70,000 events. The source types of these events are primarily classified into four distinct categories: earthquakes, explosions, surface events, and noise, with other categories only having a few samples. PNSN only locates the sources of earthquakes and explosions, which we show in Figure~\ref{fig:figure1}. PNSN analysts only identify the surface events at one or two stations; thus, their locations are assumed to be close to the seismic station that records them, as shown in Figure~\ref{fig:figure1}. The classes are not balanced (see Fig. S1): 90\% of the labeled seismic records are earthquakes, which is expected given the PNSN's mandate to monitor earthquakes. The noise class is artificially generated by \citep{ni2023curated} to provide sufficient examples and was verified using the transfer-learned earthquake transformer model \citep{mousavi2020earthquake} to ensure that it does not contain earthquake waveforms. There are over 8,000 examples of surface events and 15,000 examples of explosions. Note that this dataset was manually classified by PNSN analysts. Through our preliminary analysis, we found that a small portion of the data may have been mislabeled. 

We refer to traces as the time series of three-component ground motions, which include either a single, vertical Z component (and the other two channels filled with zeros) for the PNSN short-period instruments or the three-component seismograms of individual seismic stations. All traces were resampled to 100 Hz to homogenize the discrete time series, detrended without removing the instrument response, and stored as a three-component NumPy array. In the following, we describe the characteristics of each class.

\subsection{Earthquakes}
\begin{figure*}
\includegraphics[width=\textwidth]{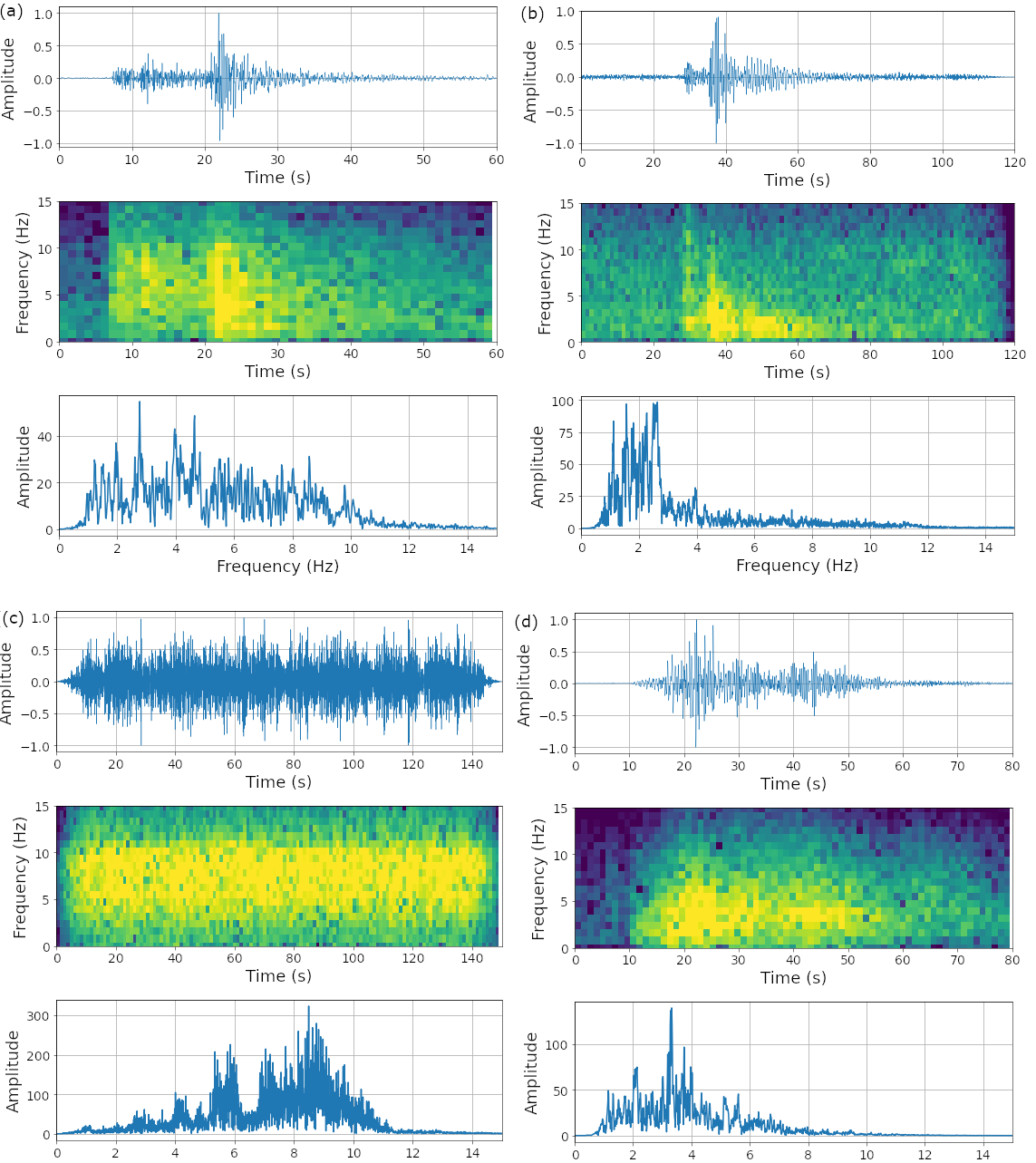}
\caption{{\bf Example of a processed single-component waveform, its corresponding spectrogram, and Fourier spectrum} of (a) earthquake, (b) explosion, (c) noise, and (d) surface event.}
\label{fig:figure3}
\end{figure*}

Earthquakes within the dataset are those whose information is sent to the ANSS Comprehensive Earthquake Catalog (ComCat) from the PNSN. \citet{ni2023curated} collected the waveform data and associated attributes in a metadata table that encompasses event-related attributes, such as source location, depth, local or duration magnitudes, and station-specific terms, including P and S picks, which PNSN analysts have generated. Their magnitudes span from -0.3 to 5 (local magnitude for most events after 2014, duration/coda magnitude for earlier events \cite{ni2023curated}); these magnitude ranges are similar to those from explosions. The PNW subduction zone hosts a diverse range of earthquake sources, including shallow crustal events, deeper intraslab earthquakes that extend to depths of ~100 km, and volcano-tectonic earthquakes. The depth distribution is bimodal: while the majority of earthquakes occur at shallow depths, a secondary concentration is found between 30–50 km, indicating the presence of intraslab seismicity. The corresponding seismic waveforms exhibit distinct P and S arrivals (Fig.~\ref{fig:figure2}), characterized by an impulsive onset and relatively higher frequencies exceeding 5 Hz (Fig.~\ref{fig:figure3}). The duration of these waveforms mostly varies between 10 and 30 seconds (Fig.~\ref{fig:figure3}). For each event sent to the ANSS ComCat catalog, \citet{ni2023curated} selected events with both P and S picks, which pre-selects for high-quality data. Each waveform in the curated dataset spans 150 seconds, sampled 50 seconds before and 100 seconds after the source origin time.

\subsection{Explosions}
PNSN analysts classify events similar to shallow quarry blasts and those occurring near recognized quarry blast sites as ``probable explosions," ``shots," or simply ``explosions." While these mechanisms differ, they are collectively categorized as "explosions." These events are characterized by a prolonged coda (Fig.~\ref{fig:figure2}) and relatively lower and monochromatic frequency content, with dominant frequencies typically falling within the range of 1 to 3 Hz  (Fig.~\ref{fig:figure3})

% Additional PX from ANSS for testing and generalizability. 

\subsection{Surface Events} \label{subsec:data_aug}
Surface events are identified at the PNSN near volcanoes, seemingly as emergent ground motions, and categorized by PNSN analysts as "surface events". This category typically includes a variety of mass-movement sources likely associated with rockfalls and avalanches, although some may also resemble low-frequency volcanic earthquakes or glacier-related activity, depending on the setting. First arrivals are typically picked at one or two stations per event and stored in the ANSS Quake Monitoring System (AQMS) database \citep{renate2020open}. Given the unknown source origin time of the surface events, waveforms are sampled from 70 seconds before the first arrival pick, as designated by PNSN analysts, and extend to 110 seconds post-P-wave pick to accommodate potentially longer-duration events. The waveforms of these events exhibit a wide range of characteristics, lasting from 20 to several minutes for the longest, but rare, debris flows (Fig.~\ref{fig:figure2}). The waveforms are less broadband than earthquake waveforms, typically falling within the range of 1 to 15 Hz, and they often feature emergent onset (Fig.~\ref{fig:figure3}). While the origin of these events is not confirmed, the PNSN analysts have mostly labeled these events at stations near volcanoes. The most active places for such labeling are Mt St Helens and Mt Rainier. A thorough characterization of these events, including their origin and mechanism, is worthy of investigation and will not be addressed in this paper. In the curated dataset, surface events were initially underrepresented, with waveform data available for only about 5,200 events (or 8,912 traces). This imbalance has hindered model performance in our preliminary exploration, limiting its ability to learn distinctive characteristics of surface events. To overcome this, we expanded the dataset by incorporating three-component waveform data from additional stations located within 30 km of each event. Although phases were not always picked and reported on these nearby stations, they often recorded strong signals due to their proximity to surface sources. This augmentation added 6,495 new three-component traces, increasing the total count to 15,407 traces and thereby enhancing the dataset's diversity, while improving the model’s capacity to generalize surface event patterns.

To further evaluate the robustness of our model as a potential surface event detector, we tested it on the 245 events from the Exotic Seismic Event Catalog hosted by Earthscope (\cite{bahavar2019exotic}, \url{https://ds.iris.edu/spud/esec}, last accessed 12/2/2024), which includes surface events from around the world that were verified by methods other than seismic. These events are mostly categorized into four source types: rock and debris avalanches, rock, debris, and ice falls, debris flows/lahars, and snow avalanches. For each event, we extracted 270 seconds of three-component data (from 70 seconds before the start time to 200 seconds after) from all stations within a 100 km radius, as exceeding this radius resulted in poorer performance due to noise levels and waveform dispersion, and we required a signal-to-noise ratio (SNR) greater than 10. Here, the SNR was computed by the ratio of means of the absolute value of the signal window (start time +30 s) over the noise window (start time - 50 s, start time -20 s)

\subsection{Noise}
The “noise” class includes 150-s waveforms extracted immediately preceding the P wave of a ComCat event. These recordings come from the curated dataset of \citet{ni2023curated}, where automated screening with a retrained Earthquake Transformer was applied to exclude hidden seismic events, resulting in very few cases identified. Given the dataset’s large size (>50,000 traces), comprehensive visual inspection of all noise records is impractical, but the deep learning picker used in the curation has demonstrated near 100\% accuracy on benchmark datasets, providing high confidence that the noise class contains minimal contamination. Nonetheless, we acknowledge the possibility that a small number of unusual or unpicked events may remain. Some of the noise recordings are characterized by numerous impulsive arrivals and non-typical impulsive earthquake signals, exhibiting a substantial amount of high-frequency content with peak frequencies typically falling within the range of 6 to 10 Hz. This distinctive waveform signature distinguishes them from typical seismic signals associated with earthquakes or explosions, further underlining their classification as ambient noise within the dataset.

\subsection{Training, Validation, and Test Data}\label{sec:train_data}

We designed separate training and validation datasets for classical machine learning (CML) and deep learning (DL) models, while using a shared testing dataset to ensure a fair comparison across approaches. Based on preliminary experiments, we observed that CML models performed best on single-component (vertical) waveforms, whereas DL models consistently outperformed when trained on three-component (3C) data. The superiority of 3C data for DL models likely reflects the importance of P/S energy ratios for distinguishing between earthquakes and explosions \cite[e.g.,][]{kong2022combining}. However, because the curated dataset contained relatively few 3C traces, we supplemented surface-event data with additional 3C recordings as described in Section~\ref{subsec:data_aug}.  

% \subsubsection{Common Test Dataset}
To evaluate models on an equal basis, we created a {\bf common test dataset} composed entirely of three-component (3C) traces. We randomly selected 10,000 traces per class. For earthquakes, explosions, and noise, the curated dataset provided sufficient 3C traces. As mentioned in Section \ref{subsec:data_aug}, for surface events, we augmented the curated dataset with 6,500 additional 3C traces from nearby stations, yielding a sufficient pool for balanced sampling. From the 10,000 traces available for each class, we randomly split the data into training, validation, and testing subsets using an 80:20 ratio. This produced 2,000 traces per class for testing (total of 8,000 traces) and reserved the remaining 8,000 per class (total of 32,000) for DL training and validation. We ensured that no event data was split between the training and testing datasets by verifying that no event identifiers were found in both the testing and the training/validation datasets.

% \subsubsection{CML Training and Validation Datasets}
We generated the {\bf CML training/validation datasets} using all events in the curated dataset, excluding the common test dataset. The training data set uses randomly sampled 6,000 traces per class. The validation data set uses the remaining 2,000 traces that do not share event ID from the training data set. If fewer than 2,000 traces were available for a class, we sampled with replacement. To account for variability, we repeated the validation sampling process 50 times with different random seeds and averaged the results across iterations.  

% \subsubsection{DL Training and Validation Datasets}
For the {\bf DL training and validation data sets}, we used the remaining 8,000 3C traces per class that remained after the common test data were set aside, and split them into 6,000 (training) and 2,000 (testing) sets, also ensuring that event ID did not leak between both subsets.

\subsection{Network Testing Dataset}\label{sec:cont_data}

To evaluate model performance in routine network operations, we generate a \textit{network testing dataset} that differs in several key ways from the curated dataset. Unlike the curated dataset, which is balanced and based on carefully reviewed analyst picks, the network dataset reflects the realities of day-to-day operations: incomplete analyst picks, class imbalance, and heterogeneous station coverage. This dataset was designed to test the robustness of models beyond controlled conditions.  

We selected the most recent events within the PNSN authoritative boundary, reviewed by the same analyst to ensure consistency across classes. A balanced set of 10,000 events was assembled, with 3,333 earthquakes, 3,333 explosions, and 3,334 surface events; noise was excluded. For each event, waveform data from up to ten stations with the earliest picks were included. Because surface events often lack locations and multi-station coverage, only a small fraction (n = 187) were located at more than one station. To address this, most surface events were supplemented with up to nine nearby stations that have historically recorded such events, typically within a 40 km radius. One more distant station, UW.JCW (>100 km, near Mt. Baker), was also included because of its frequent surface-event detections.  

The resulting dataset captures the natural variability of routine operations. Earthquakes generally had magnitudes between 0 and 2, epicentral distances below 50 km, and SNR values ranging from 0 to over 20 (capped at 20 to avoid skewing). Explosions showed similar magnitudes and SNR ranges but occurred at slightly larger distances, clustering around 50 km. Surface events lacked magnitude estimates but were typically recorded at distances of less than 25 km, with a few extending to $\sim$ 150 km. These long-distance cases may reflect either misclassified artifacts or genuine surface events at Mt. Baker, where volcano–station separations are larger (see Supplementary Fig. S13). By including these variations in class balance, SNR, and station coverage, the network testing dataset provides a realistic benchmark for assessing model performance in operational settings.

\subsection{Generalization Datasets}\label{sec:generalization_data}

An important property for models is to generalize beyond the training datasets. Because training, validation, and testing data sets were compiled for the PNW region, we generated three additional datasets: (1) the Exotic Seismic Event Catalog (ESEC), which provides verified global surface events \citep{bahavar2019exotic}, (2) a near-field explosion dataset, designed to test distance effects, and (3) incrementally expanded training datasets that incorporate additional surface events and near-field explosions. Together, these datasets allowed us to probe the limits of model generalization and to iteratively refine the training set when systematic misclassifications were observed.

\subsubsection{Exotic Seismic Event Catalog testing dataset}\label{subsec:esec_data}
The Exotic Seismic Event Catalog (ESEC) \citet{bahavar2019exotic} is an expanding global database of non-tectonic seismic events that have been verified through independent observational means (e.g., visual, remote sensing etc). These events include landslides, rockfalls, debris flows, snow avalanches, and other environmental phenomena. While the catalog spans worldwide, the majority of events are concentrated in North America and Europe. Each entry provides detailed metadata, including event origin time, geographic coordinates, and event type.

To assess the generalization capabilities of our classifier beyond the PNW, we retrieved waveforms for 245 ESEC events, selecting stations within 150 km of each event epicenter to ensure SNR and wide coverage. Since our previous retrieval, 75 new events had been added to the catalog. We included waveform data for these newly cataloged events as an additional test set, allowing us to further evaluate the model’s robustness in classifying diverse surface events across multiple regions and signal characteristics.

\subsubsection{Near-Field Explosion Test Dataset} \label{subsec:near_field_explosion_data}

We report that model performance when testing models on the ESEC dataset indicated that surface events were often misclassified as explosions at larger epicentral distances. To test whether this confusion was due to distance-dependent signal characteristics, we constructed a dedicated near-field explosion test set (0–50 km). We downloaded approximately 10,000 waveforms from $\sim$1,100 explosion events within the PNSN authoritative boundaries and saved in the ANSS catalog, restricting the data to 2023–2025 to ensure no overlap with the training set. This dataset allowed us to directly test whether models trained primarily on far-field explosion signals could generalize to near-source recordings.

\subsubsection{Incrementally Improved Training Data Sets} \label{subsec:increment_data}
The systematic misclassifications observed when testing on the ESEC and near-field datasets motivated us to expand the curated training dataset and improve generalization iteratively. We therefore developed two successive versions of the training set. Version 2 enhanced robustness to background noise and variability in explosion data by adding randomly sampled noise-only waveforms to the curated training set, effectively doubling the dataset size to $\sim$12,000 traces per class. Version 3 further expanded the dataset by incorporating two targeted sources of additional data. First, we added 1,866 surface event waveforms from the Exotic Seismic Event Catalog (ESEC), excluding lahars, debris flows, and events previously misclassified as earthquakes due to lower SNRs. Second, we included 2,502 explosion waveforms recorded at near-field distances (<50 km) from the ANSS catalog within the PNSN authoritative boundary. These additions directly addressed patterns observed during earlier testing: models trained solely on the curated dataset often confused surface events with explosions at larger distances, and some true explosions were misclassified as surface events. By selectively enriching the training data with representative surface events and near-field explosions, Version 3 aimed to reduce these confusions and expose the models to a broader diversity of event types and source characteristics.

\section{Methods}
\begin{figure*}
\includegraphics[width=\textwidth]{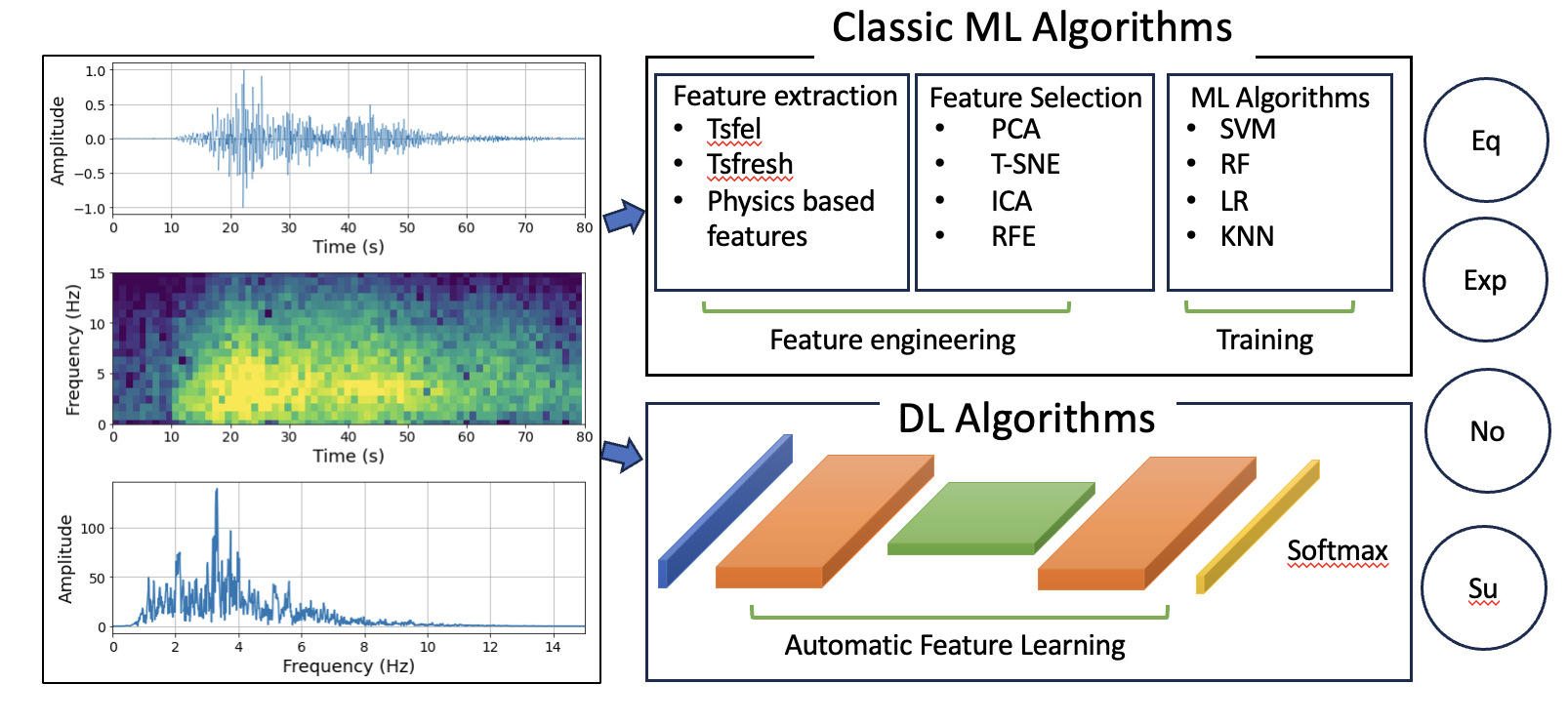}
\caption{{\bf Two main approaches to supervised machine learning in event discrimination:} Classic Machine Learning (CML) requires feature engineering before classification. Deep Learning (DL) encompasses feature extraction and classification within a single optimization framework. Input data may be the raw time series, the Fourier Amplitude Spectrum, or the short-time Fourier amplitude spectrum (i.e., spectrograms). Features are either extracted and selected or transformed before CML classification, or they are solved in a single network using deep learning (DL). Finally, the four classes are predicted: Eq (earthquake), Exp (explosion), No (noise), and Su (surface events).}
\label{fig:figure4}
\end{figure*}

We explore the two main branches of machine learning in classification, which we summarize in Figure~\ref{fig:figure4}. CML algorithms utilize engineered features \citep{jordan2015machine} that may be automatically generated using time series toolboxes such as tsfresh \citep{christ2018time} and Tsfel \citep{barandas2020Tsfel, kharita2024discrimination}, or chosen based on physical models \citep[e.g.,][]{chmiel2021machine}. The DL approach learns feature representations as part of the optimization \citep{zheng2018feature, kong2021deep}. In addition to performance and robustness, we will consider the computational cost of deploying either strategy as part of our evaluation. 

\subsection{Classic Machine Learning}

\subsubsection{Data Processing}
% Features engineering, types of features, how to extract them, how long does it take
Raw seismic data are typically high-dimensional, making them unsuitable for direct use in conventional machine learning (ML) algorithms. Feature engineering—the process of creating, transforming, and selecting new features from raw data—is therefore critical for reducing dimensionality and improving model performance of CML algorithms. The goal is to extract essential information that enhances the model’s ability to identify patterns and make accurate predictions \citep{zheng2018feature}. 

In this study, we extracted a comprehensive set of features commonly employed in ML applications for seismological analysis. Drawing on both established time-series toolboxes and prior seismological studies, we grouped features into the following categories:

\begin{enumerate}
\item \textbf{TSFEL features}: Extracted using the \texttt{TSFEL} Python library \citep{barandas2020Tsfel}, consisting of 390 features calculated from time, Fourier, and wavelet domains. TSFEL categorizes these features into statistical, temporal, and spectral domains; we did not use the recently introduced fractal domain. We selected TSFEL for its simplicity and broad coverage of features that have proven effective in prior seismological studies \citep[e.g.,][]{kharita2024discrimination, malfante2018automatic}.  

\item \textbf{Physical features}: Designed based on seismological observations and physical models of mass movements, these features capture characteristics that distinguish slope failures (surface events) from earthquakes \citep{hibert2017automatic, Provost2017, maggi2017implementation, hibert2014automated, domel2023event, kharita2024discrimination, huynh2025real}. Examples include the ratio of ascending to descending time, which differentiates impulsive fault sources (peaks at onset) from granular flows such as landslides (peaks mid-event) \citep[e.g.,][]{allstadt2018seismic, hibert2011slope}. Other features include dominant and centroid frequencies, which provide information on source type, and kurtosis and skewness across frequency bands, which indicate impulsiveness. We do not include other physics-based features used in the explosion P/S ratio \citep{kong2022combining} and various magnitude estimates \citep{Koper24} because these are not calculated for surface events.

\item \textbf{ScatNet features}: Derived from scattering convolutional neural networks \citep{anden2014deep}, these higher-order wavelet transforms have recently proven effective in distinguishing seismic sources \citep{seydoux2020clustering, moreau2022analysis, kopfli2024examining, steinmann2023machine}. Scattering transforms provide translation-invariant, noise-robust features by convolving input waveforms with wavelets across scales, forming first- and second-order scattering coefficients. We selected wavelet parameters following the guidelines in \citet{seydoux2020clustering}. Details of feature extraction are provided in the Supplementary Material Text S1.

\item \textbf{Manual features}: To account for anthropogenic patterns relevant to explosions, we added temporal descriptors such as hour of day (local time), day of week, and month of year. These features reflect human activity schedules and complement the physically motivated features.
\end{enumerate}

For the CML approach, we extracted features from vertical-component waveforms only, thereby maximizing the dataset size, as many PNSN stations record only a single component. After feature extraction, we performed standard data cleaning, removing samples with NaNs, Inf values, or identical constant values. We further reduced dimensionality by removing highly correlated features (Pearson correlation coefficient $>0.95$). Finally, we applied a threshold-based filter to remove outliers, discarding any samples where a feature exceeded five standard deviations from the mean.  

To investigate the impact of time windowing and pre-filtering on extracted features and classification performance, we constructed six feature sets (M1–M6) with varying window lengths and frequency bands. For earthquakes and explosions, windows were aligned relative to the P arrival, while for surface events, they were aligned to the analyst-defined first arrival in the curated dataset. Noise windows were extracted using the same time spans and frequency bands. The models are defined as follows:

\begin{itemize}
 \item M1: 40-second window (P$-10$ to P$+30$ s), filtered 1–10 Hz
 \item M2: 40-second window (P$-10$ to P$+30$ s), filtered 0.5–15 Hz
 \item M3: 110-second window (P$-10$ to P$+100$ s), filtered 1–10 Hz
 \item M4: 110-second window (P$-10$ to P$+100$ s), filtered 0.5–15 Hz
 \item M5: 150-second window (P$-50$ to P$+100$ s), filtered 1–10 Hz
 \item M6: 150-second window (P$-50$ to P$+100$ s), filtered 0.5–15 Hz
\end{itemize}

\subsubsection{CML Architectures}
We systematically compared seven widely used machine learning algorithms for seismic event classification: Logistic Regression (LR) \citep{hosmer2013applied}, Multi-Layer Perceptron (MLP), Support Vector Classifier (SVC) \citep{hearst1998support}, K-Nearest Neighbors (KNN) \citep{cover1967nearest}, Random Forest (RF) \citep{breiman2001random}, XGBoost (XGB) \citep{chen2016xgboost}, and LightGBM (LGBM) \citep{ke2017lightgbm}. Our evaluation resembled an automated ML approach in scope but was implemented manually and transparently: we defined parameter ranges for each algorithm, performed initial hyperparameter searches, and directly compared their performance.  

Each algorithm was assessed using five-fold cross-validation, with the macro F1 score as the optimization metric. This stage of analysis was intended to identify which algorithm families consistently performed well on our data. Results are shown in Figures~S3–S4, with tested parameter ranges and optimal values summarized in Supplementary Text~S2. Tree-based algorithms (RF, XGB, and LGBM) consistently outperformed others. LGBM achieved the highest average macro F1 score (~89\%), followed by RF (~85\%). However, training LGBM was at least 20 times slower than RF under comparable settings, and runtime scaled steeply with larger grids. Because our study required hyperparameter tuning across 48 feature sets and multiple validation seeds, tuning was computationally exhaustive. RF offered a more practical balance: strong performance, lower computational costs in training, and the ability for interpretable feature importance estimation. In addition, RF is widely used in seismology because of its robustness, relatively simple hyperparameter space, and intuitive feature-importance measures \citep[e.g.,][]{hibert2014automated, hibert2017automatic, Provost2017, kharita2024discrimination}. For these reasons, we selected RF as the primary CML algorithm for the rest of the study.  

\subsubsection{Training and Tuning}
Choosing RF only, we carried out a second, more extensive hyperparameter optimization tailored to each of the 48 feature sets: we randomly sampled approximately 300 hyperparameter combinations from the search space, varying \texttt{n\_estimators}, \texttt{max\_depth}, \texttt{min\_samples\_split}, and \texttt{min\_samples\_leaf}. Each candidate configuration was evaluated using five-fold cross-validation, and the best-performing combination (macro F1) was selected. To control runtime, we tuned on a balanced subset of 3,000 samples per class. Across feature sets, this deeper optimization typically yielded limited additional 1–2\% improvement in macro F1 compared to the initial broad search. Detailed parameter ranges are available in the public software repository linked in the code availability section.

\subsection{Deep Learning} 
% DL algorithms are deep neural networks with multiple layers of interconnected nodes (neurons) whose parameters are tuned during optimization to minimize a cost function and correctly predict the event class based on time series. Examples of DL algorithms are 1D and 2D convolutional neural networks . 

\subsubsection{Data Processing}

For deep learning models, we prepared three-component waveform traces using a standardized preprocessing pipeline. For each trace, we extracted a 100-second window, with the window start chosen randomly between 5 and 20 seconds before the analyst's pick time. All waveform traces were preprocessed before model training and evaluation. Each trace was first linearly detrended to remove baseline offsets and then tapered with a 1\% cosine taper to minimize edge effects introduced during filtering. We applied a bandpass filter between 1–20 Hz to suppress long-period noise as well as high-frequency artifacts, ensuring the retention of the frequency content most relevant for seismic event discrimination. Finally, each trace was normalized by its standard deviation, following common practice in deep learning applications for seismic data, so that all channels contribute comparably during training. To ensure data quality, we only retained traces with an SNR greater than 1, computed following the same procedure described by \citet{ni2023curated}. In addition to raw waveform inputs, we generated spectrogram representations of each three-component trace. Spectrograms provide a joint time–frequency characterization of the signals, which has been shown to improve classification performance in many deep learning applications. We implemented a PyTorch-based spectrogram computation function (\texttt{compute\_spectrogram}), which transforms each input batch of waveforms $(B, C, T)$ into power spectral density representations $(B, C, F, T_{\mathrm{spec}})$. Waveforms were segmented into overlapping windows using a Hann taper, with each segment set to 256 samples (\texttt{nperseg = 256}) and 50\% overlap. Each segment was then mean-centered, tapered, and transformed into the frequency domain using the real-valued FFT. Power spectral density (PSD) estimates were obtained by normalizing the squared magnitudes of the Fourier coefficients by the window power and sampling rate. To ensure energy conservation under the Parseval convention, one-sided spectra were produced by doubling all frequencies except the DC and Nyquist components. Finally, frequency and time axes were derived from the FFT length and hop size, providing consistent alignment across all traces. The resulting spectrograms preserve three channels per trace (E, N, Z), while adding frequency and time dimensions, producing an input well-suited for convolutional neural networks. By combining both waveform- and spectrogram-based representations, we ensured that the deep learning models had access to both temporal and spectral features of the seismic signals.

\subsubsection{Model Architectures}
In recent years, deep learning methods, particularly Convolutional Neural Networks \citep[CNN,][]{lecun2015deep}, have emerged as powerful algorithms for seismic event classification, leveraging their ability to extract hierarchical features from raw waveform data automatically. CNNs excel in learning complex patterns directly from the data, making them particularly effective for distinguishing between events such as earthquakes, explosions, surface events, and noise. In recent studies, CNNs have been successfully applied to process seismic signals, demonstrating superior performance over conventional methods in both accuracy and scalability \citep[e.g.,][]{mousavi2022deep}. The multi-layered architecture of CNNs allows them to capture both local and global features in seismic waveforms, making them well-suited for detecting subtle variations in seismic signatures. Moreover, CNNs can be extended to incorporate multi-channel inputs, such as combining signals from different seismic components, which further enhances their classification capability \citep[e.g.,][]{mousavi2022deep, ross2018p, perol2018convolutional,linville2019deep}. As a result, CNNs have become a valuable tool in seismic monitoring, aiding in the real-time detection of events and enhancing the accuracy of seismic discrimination systems.

Designing an optimal CNN architecture is a complex task, primarily due to the many hyperparameters that must be tuned for best performance. These hyperparameters include the number and size of filters, the configuration of the convolutional, fully connected, dropout, and pooling layers, as well as the choice of activation functions. The search for an ideal hyperparameter combination is computationally intensive and far more demanding than traditional machine learning approaches, often requiring heuristic optimization methods, such as random search or grid search, to explore the vast parameter space \citep[e.g., Network Architecture Search - NAS, ][]{elsken2019neural}. Alternative methods may reuse the architecture of foundation models (e.g., VG16).

Each study employs a specific model architecture, topology, and dataset, making intercomparison challenging. To remediate this, without undergoing a full NAS, we propose representing diversity in architecture by utilizing two canonical architectures and two types of inputs: time series and spectrograms. We design a wide/fat shallow network (SeismicCNN) and a skinny deep network (QuakeXNet), which we illustrate in Fig. S5. For each architecture, we have a (1D) and a (2D) version, where (1D) refers to taking time series input data and (2D) refers to taking spectrograms as inputs. Spectrograms are a transformation of the raw data or another form of feature space. The time series has a dimension of 3x5001 points, and the spectrograms have a dimension of (3x129x38). Each architecture is designed to address the unique characteristics of seismic waveform data, aiming to improve classification performance.

% SeismicCNN architecture
The SeismicCNN architecture consists of two convolutional layers, each followed by a batch normalization layer and a max pooling layer. The architecture begins with a 1D convolutional layer that extracts features from the seismic waveforms. The first layer applies 32 filters, each with a kernel size of five, followed by a batch normalization and ReLU activation. The second convolutional layer expands to 64 filters, again with a kernel size of five, followed by batch normalization and ReLU activation. Both convolutional layers are followed by max-pooling layers with a kernel size of two, which reduces the temporal dimension and focuses on key features. Dropout layers with a rate of 0.2 are integrated after each pooling operation to prevent overfitting during training. The final layers consist of fully connected layers that output the class probabilities for various seismic events. SeismicCNN (2D) has the same architecture but instead uses spectrograms as input and 2D convolutions. However, the size of the models has a significant difference: SeismicCNN (1D) has 10,227,340 parameters, whereas SeismicCNN (2D) has 1,986,572 parameters, primarily due to the difference in input size.

% QuakeXNet architecture
The QuakeXNet (1D) architecture consists of seven convolutional layers, each followed by batch normalization layers and two max pooling layers. The architecture begins with sequential 1D convolutional layers, each followed by batch normalization and ReLU activation. The first convolutional layer uses eight filters with a kernel size of nine and padding to preserve input dimensions, while subsequent layers gradually increase the number of filters to 64 and alternate between stride lengths of one and two. This progressive increase in the number of filters is thought to allow the model to capture more complex features at deeper layers. A max pooling operation is applied after every second convolutional layer to reduce the temporal resolution, enabling the network to focus on the most salient features, and is a form of blur pooling layer \citep{zhang2019making} and has been utilized in denoising CNNs \citep[e.g.,][]{yin2022multitask}. The classifier takes the flattened output of features from the convolutional layers and passes it through a fully connected layer with 128 neurons, followed by batch normalization and ReLU activation, and another fully connected layer that outputs class logits. Dropout layers with a rate of 0.2 are inserted after pooling layers and fully connected layers to mitigate overfitting. The QuakeXNet (2D) architecture, adapted for 2D seismic spectrogram inputs, mirrors the QuakeXNet (1D) structure but uses 2D convolutional layers. QuakeXNet (1D) has 657,716 parameters, whereas QuakeXNet (2D) has 70,708 parameters.

\subsubsection{Training and Tuning}
% training these NNs
For training, we used the Adam optimizer,  an initial learning rate of 0.001, and a cross-entropy loss. We train on a single NVIDIA RTX3090 24GB RAM. We trained the models with a batch size of 128 for up to 100 epochs, implementing early stopping if validation performance did not improve within 30 epochs. This approach allowed us to fine-tune the model’s ability to generalize while avoiding overfitting on the training set. We assess the performance of each model by comparing the training and validation losses, along with the validation accuracy, as the number of epochs increases.

%% reports on training these NNs
The training loss of the SeismicCNN (1D) architecture showed a smooth and consistent decline over the epochs, indicating effective learning during training. In contrast, the validation loss exhibited a more irregular decay with the number of epochs, characterized by an initial increase followed by a gradual decrease. Training SeismicCNN achieved a peak validation accuracy of 89\%  (Fig. S6). In comparison, QuakeXNet (1D) demonstrated improved stability in the training, with the training loss steadily decreasing from 0.75 to 0.15 and the validation loss dropping from 0.9 to 0.45. The validation accuracy improved significantly, rising from 60\% to 92\%, indicating that this architecture was highly effective for the given task (Fig. S6). 

Given the limitations observed with the 1D models, we shifted our focus to 2D input architectures. The SeismicCNN (2D) architecture demonstrated that both training and validation losses decreased smoothly, with the training loss falling from 0.75 to 0.15 and the validation loss dropping from 0.75 to 0.25. The accuracy improved from 75\% to 94\%, highlighting the model's effectiveness in discriminating between classes. Similarly, the QuakeXNet (2D) architecture showed consistent improvement, with both training and validation losses decreasing from 0.9 to 0.3. The accuracy increased steadily from 78\% to 92\%, although the validation loss was slightly lower than the training loss. Overall, our findings suggest that transitioning from 1D to 2D architectures improves the model's ability to distinguish between various seismic event classes (Figs. S7 and S8).

\subsection{Model Deployment Workflows}

To explore how our models could be applied in operational and research settings, we designed several deployment workflows. These workflows span real-time and retrospective use cases, ranging from integration with existing software ecosystems to large-scale cloud-ready pipelines. The following subsections describe the implementation details of each workflow.

\subsubsection{Integration with SeisBench}
We first implemented our deep learning classifiers in the {\tt seisbench} ecosystem \citep{Woollam22}, enabling seamless integration into catalog-building workflows. The models are compatible with the {\tt model.classify} and {\tt model.annotate} functions, which return class labels, timestamps, and probability traces for use in real-time or retrospective analysis.

\begin{lstlisting}[caption=Read an obspy stream and classify or annotate with seisbench, label=code, language=Python]

# importing the dependencies
import seisbench.models as sbm 
import obspy
# load the model
model = sbm.QuakeXNet.from_pretrained()
model.eval()  # Set to evaluation mode
# reading the data
z = obspy.read("file.mseed")
# classify the data
results = model.classify(z)  
# anotate the data
annotations = model.annotate(z)
\end{lstlisting}

\subsubsection{Deployment on the network-testing dataset (retrospective)}
To evaluate model performance under controlled conditions, we applied our classifiers to the network test dataset. For each event, we retrieved waveforms from the 10 nearest stations and extracted 141-second windows (30~s before to 111~s after origin), resampled to 50 Hz. We then ran two DL models (QuakeXNet and SeismicCNN) and two ML models (M2\_30 and M2\_110) with a five-second stride. Station-level probabilities were averaged and filtered using quality-based criteria (SNR thresholds, probability thresholds, probability-distance thresholds). Event-level predictions were obtained by majority vote across high-confidence stations. To assess robustness, we conducted a grid search over threshold values.

\subsubsection{Deployment on continuous data (cloud-ready workflow)}
Finally, to test how to run this continuously and deploy these on cloud data archives \citep[e.g.,][]{Ni2025a,Ni2025b}, we developed a pipeline to scan daily waveform archives stored on Amazon Web Services S3. Day-long miniseed files were read, relevant channels (e.g., HH*, BH*) extracted, and data preprocessed (filtering, trimming, spectrogram computation). Inference was performed using 100-second windows with a 20-second stride. Probability traces were smoothed with a moving average (window size = 5). Detections were initiated when the smoothed probability exceeded 0.15 and ended when it dropped below, with valid detections requiring a maximum probability above 0.5. Each detection was assigned the class with the highest peak probability, and outputs included labels and timestamps.

\section{Model Performance Results}

We structured our performance evaluation in stages, progressively narrowing down the best candidate models and testing their robustness across increasingly realistic and challenging datasets. First, we benchmarked all classical machine learning (CML) and deep learning (DL) models on a balanced test set derived from the curated dataset (Section~\ref{sec:train_data}). This initial evaluation allowed us to compare feature-based CML approaches against end-to-end DL classifiers under controlled conditions. Next, we selected the top-performing models—two CML models and two DL models—and tested them on the network testing dataset (Section~\ref{sec:cont_data}), which simulates real-world network conditions with class imbalance, varying SNR, and heterogeneous station coverage. These experiments assessed how well models trained on curated data generalize to routine network operations. We then evaluated the best DL models on two additional independent datasets designed to probe generalization: the Exotic Seismic Event Catalog (ESEC) dataset and a near-field explosion dataset. These datasets revealed systematic misclassifications, particularly confusion between surface events and explosions, motivating the need for further improvements. Finally, to address these shortcomings, we iteratively retrained the DL models with augmented training data, creating Version~2 and Version~3 datasets. Each version incorporated additional representative traces (e.g., noise-only records, surface events from ESEC, near-field explosions), progressively improving robustness. The Version~3 model emerged as the best-performing classifier, striking the best balance between accuracy, robustness, and computational efficiency.

\subsection{Evaluation Setup}

We evaluated model performance using standard classification metrics. \textit{Precision} is the proportion of true positive predictions out of all predicted instances for a given class, while \textit{recall} (or sensitivity) is the proportion of true positives out of all actual instances of that class. At a broader level, \textit{accuracy} is the ratio of correctly predicted instances (true positives and true negatives) to the total number of instances, and the \textit{F1 score} is the harmonic mean of precision and recall, providing a balanced measure that accounts for both false positives and false negatives. F1 is particularly informative when datasets are imbalanced.  

Model outputs were evaluated at two levels. At the \textit{station level}, predictions were generated independently for each waveform trace, with the assigned class corresponding to the maximum predicted probability among the four classes (earthquake, explosion, surface event, or noise). At the \textit{event level}, predictions from all available stations were aggregated by averaging class probabilities across stations, and the event was assigned to the class with the highest mean probability. For the balanced curated test dataset, station coverage was often limited—particularly for surface events, so we report only station-level performance. For the network testing dataset, where more complete station coverage was available, both station- and event-level performance were assessed (see Section~\ref{sec:cont_data}). This evaluation framework provides the basis for the performance comparisons presented in the following sections.

\subsection{Performance on the Curated Test Dataset}

We first evaluated the performance of classical machine learning (CML) models trained on engineered features and deep learning (DL) models trained on waveform time series or spectrogram inputs. This controlled test set provided a benchmark for comparing the two approaches under balanced conditions.

\subsubsection{CML Performance}
We evaluated eight feature sets and their combinations to classify seismic traces using CML models: Scatnet features (110), Scatnet + manual features (113), Tsfel (390), Tsfel + manual (393), Physical (62), Physical + Manual (65), Physical + Tsfel (454), and Physical + Tsfel + Manual (457).

\begin{figure*}[h!]
\includegraphics[width=\textwidth]{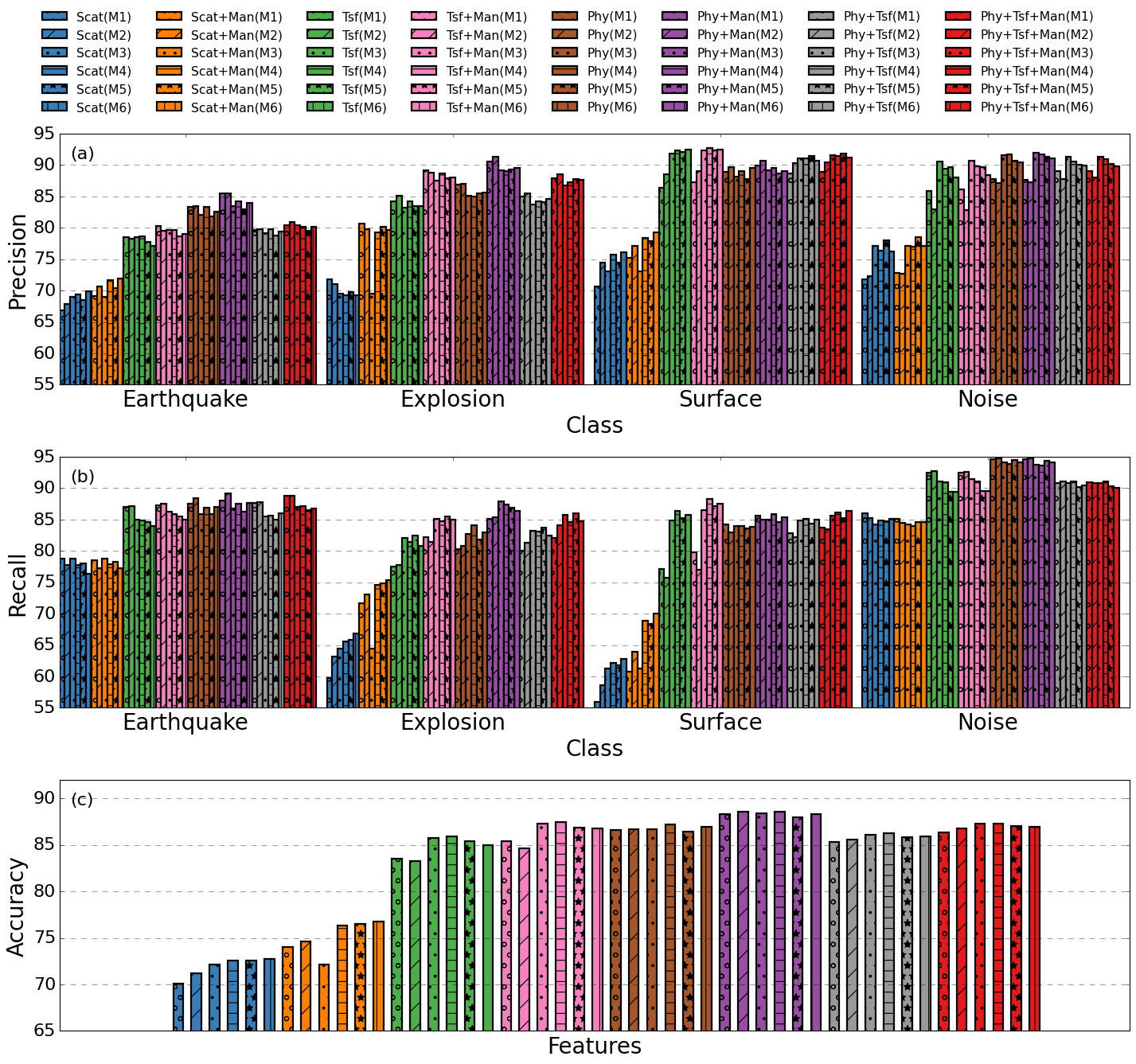}
\caption{{\bf CML Model Performance.} Precision (a), recall (b), and accuracy (c) for random forest models trained on different feature sets and waveform configurations.}
\label{fig:figure5}
\end{figure*}

\paragraph*{Precision}
Precision results shown in Figure~\ref{fig:figure5} highlight the importance of certain feature groups. Feature sets with Scatnet-derived features alone showed moderate precision, which improved when combined with Manual features. Tsfel-based features outperformed Scatnet. Combining Physical features with others consistently achieved the best precision, especially for earthquakes and explosions. Longer waveform windows (110–150 s; M3–M6) showed a slight improvement in precision for explosions and surface events compared to shorter windows (40 s; M1–M2), although the differences were generally small (on the order of 1–2\%). For the best-performing feature set (Physical + Manual), shorter 40-s windows (M1–M2) performed marginally better than longer ones, underscoring that window length did not have a consistent effect across all feature sets. Broader frequency ranges (0.5–15 Hz; M2, M4, M6) tended to yield marginally higher precision than narrower bands (1–10 Hz; M1, M3, M5), but again the gains were minor. Noise precision remained stable across all configurations, with Physical + Manual features consistently achieving the highest values overall.

\paragraph*{Recall}
Hybrid feature sets that included Physical features yielded higher recall across all classes (Fig.~\ref{fig:figure5}). Longer windows again improved recall for explosions and earthquakes, and broader frequency ranges (0.5–15 Hz) contributed to stronger performance compared to narrower filters. Noise recall was consistently high, reflecting its distinctive spectral content. For physical features, a negligible difference in performance was observed.

\paragraph*{Accuracy}
Accuracy values ranged from moderate (70–73\% for Scatnet features) to strong (86–88\% for Physical-based features). Tsfel features significantly improved accuracy (83–86\%), especially when combined with Manual features (up to 87\%). Overall, feature type was the dominant factor for CML performance, followed by waveform window length, while preprocessing choices had smaller effects (Fig.~\ref{fig:figure5}). The best accuracy (88.5\%) was achieved by Physical + Manual features, underscoring the value of physically interpretable descriptors.

\subsubsection{DL Performance}

We next evaluated four DL models: SeismicCNN (1D, 2D) and QuakeXNet (1D, 2D) with waveform time series (1D) or spectrograms (2D) as input.r 

\begin{figure*}
\includegraphics[width=\textwidth]{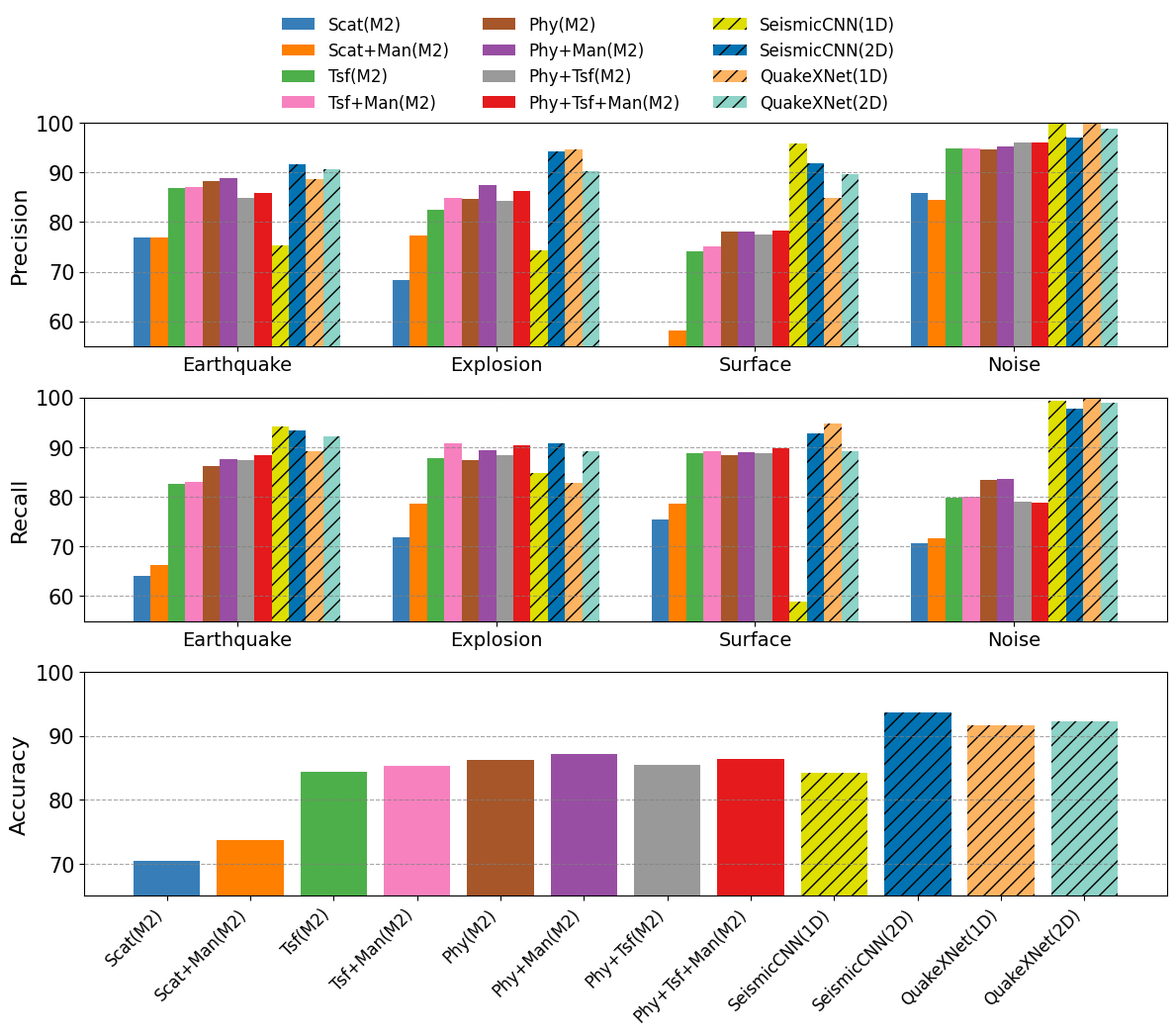}
\caption{{\bf DL and CML Model Performance.} Precision (top), recall (middle), and accuracy (bottom) on the balanced curated test dataset. Histograms are color-coded by model family.}
\label{fig:figure6}
\end{figure*}

\paragraph*{Precision}
Precision results shown in Figure~\ref{fig:figure6}a vary across models and classes. SeismicCNN (1D) achieved high precision for surface events (95\%) and noise (99\%), but struggled with earthquakes (75\%) and explosions (74\%). Using spectrograms improved results substantially: SeismicCNN (2D) reached 90–96\% precision across all classes. QuakeXNet (1D) performed well for noise (99\%) and earthquakes (88\%) but showed lower precision for surface events (84\%). QuakeXNet (2D) achieved the strongest and most balanced precision (90–98\%) across all classes.  

\paragraph*{Recall}
Recall values shown in Figure~\ref{fig:figure6}b showed similar trends. SeismicCNN (1D) recalled earthquakes (94\%) and noise (99\%) well, but misclassified many surface events (58\%). SeismicCNN (2D) delivered balanced recall across classes (90–97\%), excelling in surface event detection (95\%). QuakeXNet (1D) achieved high recall for surface events (94\%) and noise (99\%) but lagged on explosions (82\%). QuakeXNet (2D) again provided the best balance (89–99\%), with strong performance across all classes.  

\paragraph*{Accuracy}
Accuracy is shown in Figure~\ref{fig:figure6}c and further highlights the advantage of spectrogram inputs. SeismicCNN (1D) achieved 84.2\%, while SeismicCNN (2D) reached 93.7\%. QuakeXNet (1D) reached 91.6\%, and QuakeXNet (2D) 92.4\%.  

\subsubsection{Comparison of CML and DL Approaches Within Domain}

DL models, particularly spectrogram-based architectures (SeismicCNN 2D and QuakeXNet 2D), consistently outperformed CML models across all metrics (precision, recall, and accuracy). While CML models benefited from carefully engineered features, their performance plateaued below the best DL models. In contrast, DL approaches leveraged end-to-end learning to capture subtle waveform differences—especially between surface events and explosions—leading to superior generalization. Overall, these results establish DL spectrogram-based models as the strongest candidates for deployment, setting the stage for further evaluation on more challenging and realistic datasets.

\begin{table}[h!]
\centering
\begin{tabular}{|l|c|c|}
\hline
\textbf{Model} & \textbf{Accuracy} & \textbf{F1 score} \\ \hline
\textit{SeismicCNN (1D)}   & 84\% & 84\% \\ \hline
\textit{QuakeXNet (1D)}    & 92\% & 92\% \\ \hline
\textit{\textbf{SeismicCNN (2D)}}   & \textbf{94}\% & \textbf{94}\% \\ \hline
\textit{QuakeXNet (2D)}    & 92\% & 92\% \\ \hline
\textit{Phy+Man (M2)}      & 89\% & 87\% \\ \hline
\textit{Phy+Man (M4)}      & 88\% & 88\% \\ \hline
\textit{Phy+Man (M6)}      & 87\% & 89\% \\ \hline
\end{tabular}

\caption{Performance of different models on the common test sets from the curated catalogs described in Section~\ref{sec:train_data}.}
\label{tab:model_accuracy}
\end{table}

\begin{figure*}
\includegraphics[width=\textwidth]{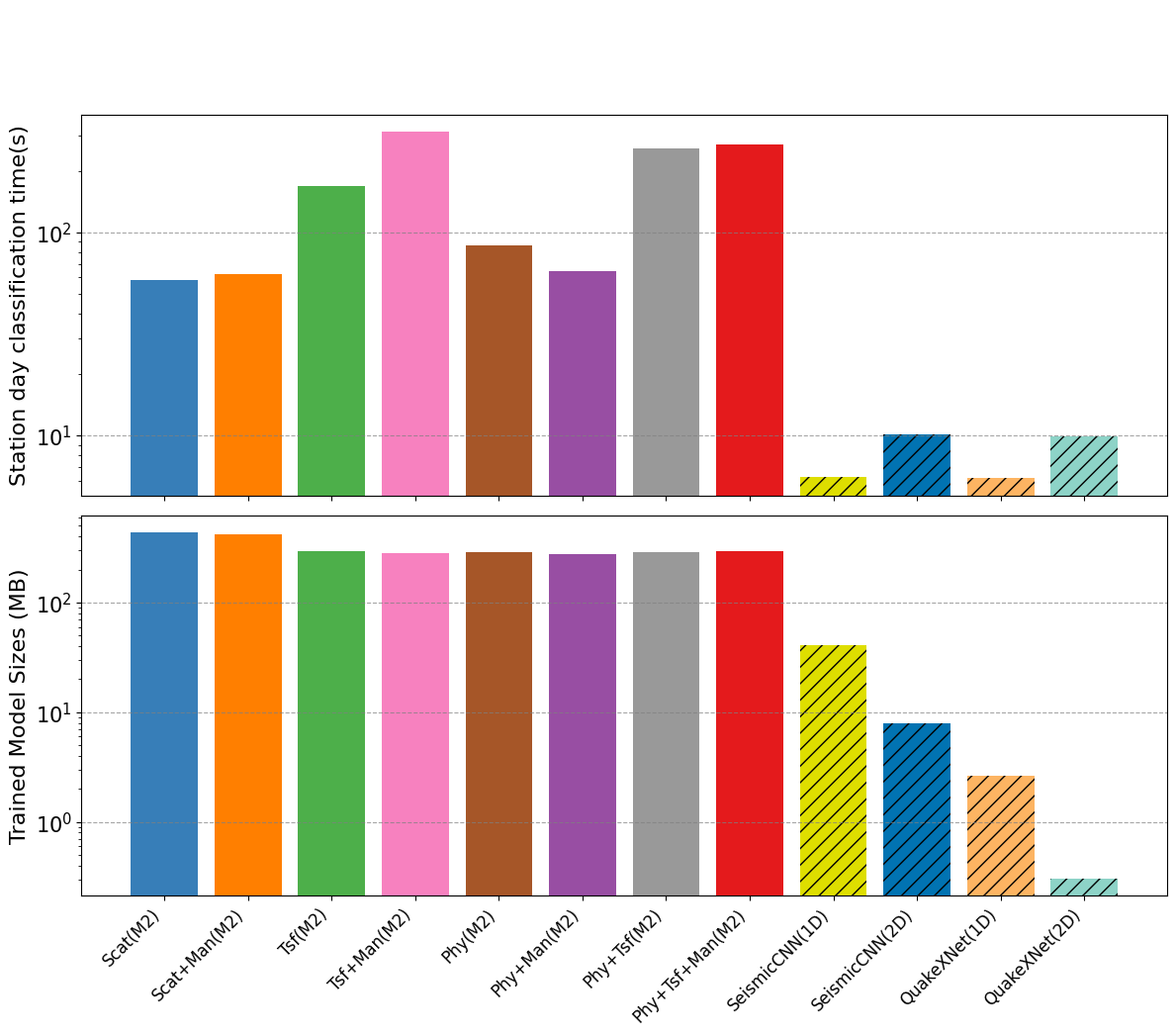}
\caption{{\bf Performance comparisons of various ML and DL models in terms of (a) station-day classification time and (b) trained model sizes }} 
\label{fig:figure7}
\end{figure*}

\subsection{Performance on the Network testing dataset}

From the balanced curated test dataset, we identified three models that achieved the best overall performance: two deep learning classifiers (SeismicCNN (2D) and QuakeXNet (2D) ) and one classical machine learning model (RF trained on Physical + Manual features with a 40-s, 0.5–15~Hz window, hereafter referred to as ML\_40). These models capture both end-to-end feature learning (DL) and engineered feature-based approaches (CML), providing complementary perspectives on event classification. In this section, we evaluate these top-performing models on the network testing dataset to assess how well they generalize under realistic operational conditions.

The two deep learning models performed robustly, with accuracy and F1 scores only slightly lower than on the curated test set (Supplementary Fig. S14). SeismicCNN (2D) achieved nearly equal performance across classes, with precision/recall/F1 of 96\%/95\%/95\% for earthquakes, 98\%/93\%/95\% for explosions, and 92\%/97\%/95\% for surface events. QuakeXNet (2D) performed comparably overall but showed a noticeable drop in precision for surface events (87\%), indicating greater confusion between surface events and explosions.  

By contrast, the ML\_40 model underperformed relative to the deep learning approaches, particularly for surface events. It correctly classified only about three-quarters of SU cases (recall = 77\%), roughly 10\% lower than the CNNs in mean F1 score. This performance gap underscores the advantage of end-to-end feature learning, which appears better suited to the heterogeneous and lower-quality records typical of routine operations.  

To further investigate the causes of misclassification, we analyzed classification probability as a function of SNR and epicentral distance (Fig. S15). For both CNNs, high-confidence predictions ($P > 0.9$) were maintained for traces with SNR > 5 out to distances of ~20 km. Confidence declined sharply for low-SNR waveforms and for distances greater than ~60 km. Surface events were most vulnerable to this decline in performance over SNR and distance, which we interpret as a greater attenuation of signals in shallow wave propagation, explaining the precision shortfall observed for QuakeXNet (2D).  

These results demonstrate that while DL models generalize well from curated to network data, their performance remains bounded by physical SNR–distance trade-offs inherent to the network geometry and noise environment, rather than by model bias.

\subsection{Performance on the Generalization Datasets}
When testing on out-of-domain datasets, however, SeismicCNN (2D) showed low performance and suggested limited generalization. On the Exotic Seismic Event Catalog (ESEC), the model frequently misclassified surface events as explosions, while on the near-field explosion dataset, it tended to confuse explosions with surface events. These results indicated that while the model was highly effective on curated PNW data, it struggled with signal characteristics not well represented in the training set.

To improve generalization, we progressively expanded the training data (subsection~\ref{subsec:increment_data}). Version 2 included noise augmentations, while Version 3 added 1,866 ESEC surface event traces and 2,502 near-field explosion traces. Version 2 improved slightly, but there was still confusion in recognizing out-of-domain surface events. Version 3 improved the classification of ESEC surface events but still struggled with near-field explosions, whereas Version 1 (curated training only) had shown the opposite trend. This highlights a trade-off in generalization: additional data helped with some out-of-domain cases but introduced new weaknesses.

Closer inspection of probability curves revealed systematic patterns: surface event probabilities tended to peak early in the waveform, while explosion probabilities peaked later, consistent with similarities in their physical sources (surface waves, explosive or detachment phases). By aggregating probabilities across entire traces and assigning labels based on the dominant class, Version 3 achieved improved event-level accuracy on both datasets. QuakeXNet (2D), when analyzed with the same procedure, generalized better than SeismicCNN (2D), suggesting that architectural differences influence out-of-domain robustness. 

Despite these improvements, a slight confusion between surface events and explosions persisted, particularly for events with emergent arrivals, extended codas, or poor signal quality. This underscores the importance of both dataset diversity and robust evaluation across global test sets.

\subsection{Overall Performance}

Across all stages of evaluation, deep learning models consistently outperformed classical machine learning approaches. On the balanced curated test dataset, spectrogram-based architectures (SeismicCNN~2D and QuakeXNet~2D) achieved the highest precision, recall, and accuracy, clearly surpassing feature-engineered CML models. On the network testing dataset, both CNNs maintained strong performance under realistic operational conditions, though QuakeXNet~2D showed a modest loss of precision for surface events compared to SeismicCNN~2D.  

Generalization tests revealed the limitations of training only on curated data sets: both CNNs struggled when applied to out-of-domain datasets, such as the Exotic Seismic Event Catalog (ESEC) and near-field explosions, frequently confusing surface events with explosions. These systematic errors motivated iterative augmentation of the training data. Version~2 incorporated noise-only records, yielding modest improvements, while Version~3 further expanded the dataset with 1,866 ESEC surface event traces and 2,502 near-field explosion traces.  

\textbf{QuakeXNet~2D Version~3 emerged as the best-performing model overall}. It provided the strongest balance of accuracy, robustness, and computational efficiency, outperforming SeismicCNN~2D on out-of-domain tests while maintaining high performance on curated and network datasets. Importantly, QuakeXNet~2D is lightweight, requiring only 70,708 parameters and $\approx$1.2~MB of memory, with a full day of continuous data processed in $\approx$9~s (Table~\ref{tab:model_performance}) at a stride of 10 s. This combination of accuracy, robustness, and efficiency makes \textbf{QuakeXNet~2D~v3} the most reliable classifier developed in this study and the most suitable candidate for deployment in real-time network operations.  

\begin{table}[ht]
\centering
\begin{tabular}{|l|r|r|r|}
\hline
\textbf{Model} & \textbf{Number of parameters} & \textbf{Memory usage (MB)} & \textbf{Inference on 1 day of 100-Hz data (s)} \\ \hline
QuakeXNet (1D) & 657,716 & 4.55 & 6.17 \\ \hline
SeismicCNN (1D) & 10,227,340 & 46.39 & 6.22 \\ \hline
SeismicCNN (2D) & 1,986,572 & 11.61 & 10.07 \\ \hline
QuakeXNet (2D) & \textbf{70,708} & \textbf{1.22} & \textbf{9.13} \\ \hline
\end{tabular}
\caption{Computational performance of selected deep learning models}
\label{tab:model_performance}
\end{table}

\section{Feature Importance}

\subsection{Feature Importance from CML}
\begin{figure*}
\includegraphics[width=\textwidth]{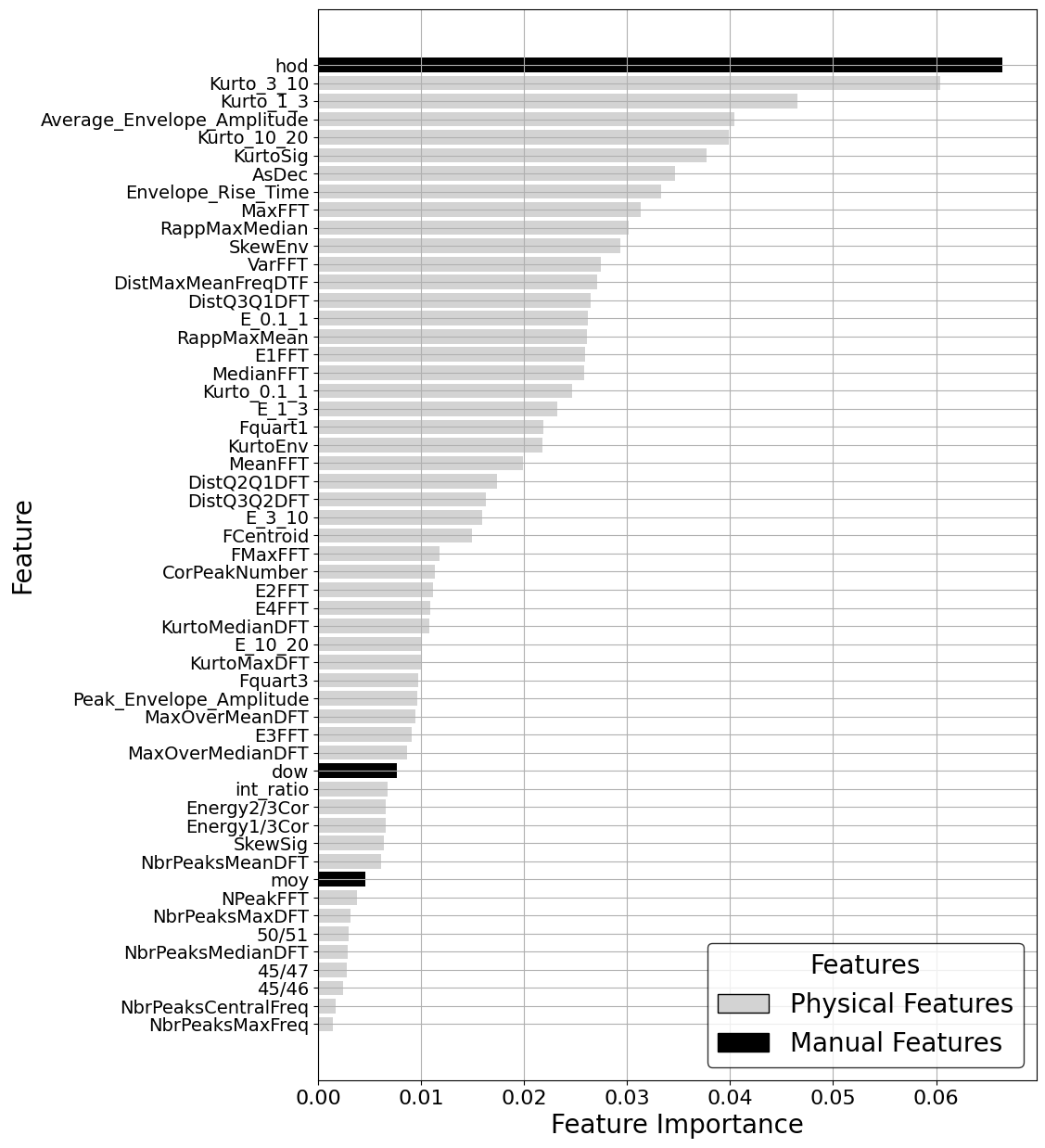}
\caption{{\bf Feature importance using model M2 and the combination of Physical and Manual features, averaged over ten iterations.}} 
\label{fig:figure8}
\end{figure*}

Focusing on the CML model that includes Tsfel, Manual, and Physical features, we now explore the importance of the feature calculated by the Random Forest algorithm and show the feature importance in Figure~\ref{fig:figure8}. In Random Forest models, the importance of a feature is typically measured by the decrease in a performance metric, such as Gini impurity or accuracy, when the feature is used to split the data in a tree. To estimate feature importance, RF models aggregate the impact of each feature across all trees. The more a feature reduces uncertainty in predictions, the higher its importance score will be. This approach has several advantages: it handles large datasets and high-dimensional spaces easily and provides a measure of feature importance without the need for feature scaling or normalization.

Figure~\ref{fig:figure8} shows the feature importance of running model M2 with a combination of Physical+manual features. The analysis of the feature set revealed that Kurtosis-based features were the most important (Fig.~\ref{fig:figure8}). Kurtosis is a statistical measure that indicates the flatness of the signal amplitude distribution compared to a normal distribution. Signals with lower kurtosis values have flatter distributions and shorter tails, indicating fewer extreme values. The feature \textit{Kurt\_3\_10} (Kurtosis in the 3-10 Hz frequency band) was found to be the most important among all time-series features, followed by \textit{KurtoSig} (kurtosis of the entire signal, filtered between 0.5-15 Hz), \textit{Kurto\_10\_20} (Kurtosis in the 10-20 Hz band, though our data was pre-filtered to 15 Hz), and \textit{Kurto\_1\_3} (kurtosis in the 1-3 Hz band). These kurtosis-related features highlight the significance of amplitude distribution in different frequency bands for classifying seismic events. When looking at histograms of the distribution of the kurtosis values over the four classes, we see that they are relatively well separated among the four classes: the noise kurtosis ranges between -0.01 and 1, the surface-event kurtosis between 1 and 20, the explosion kurtosis varies between 2 and 25, and the earthquake kurtosis varies between 6 and 60. These kurtosis-based features provided a strong separation between each class. In addition to kurtosis-based features, the manual feature ``hod" (hour of day) was found to be the most important feature. This was expected, as explosions typically occur during the day, making this feature highly informative for discriminating explosions from other events (Fig. S11). Other notable features included the ``Average envelope amplitude," which effectively shows the shaking duration and provides a good way to distinguish long-duration surface events from shorter-duration earthquakes and explosions, and ``Envelope rise time," where surface events demonstrate a slower growth. 

While Random Forest is a powerful algorithm for classification and explainability, it has some limitations. There is a bias toward high cardinality features as decision trees tend to assign higher importance to features with many unique values (high cardinality), regardless of whether they are truly informative. When features are correlated, the model tends to distribute importance across them. This can result in underestimating the importance of the more influential feature, but it is addressed here by removing highly correlated features. Finally, the importance scores may vary significantly across different runs, especially if the dataset contains noise or irrelevant features. We mitigated this by averaging the importance calculated over ten iterations.

\paragraph{Feature Selection}
To identify the minimum number of features required for comparable performance to the full set of features, we computed the performance of our model with a progressively increasing number of the most important features. We find that the 20 most important features suffice to predict with an F1 score of 89\%, which is as much as a model that includes 62 features (Fig. S9). Further, using just a single feature provides an F1-score of 60\%, whereas using just the 20 most important features provides an F1-score of 90\%. This results in a reduction of computational time by approximately 1.5, while providing similar performance (Fig. S9).

\subsection{Feature importance from DL}
% \subsubsection{Feature Importance in Deep Learning}
Difficulties in the interpretability of neural networks have been a major limitation for the broad adoption of deep learning methods in seismic discrimination. \citet{kong2022combining} was among the first to utilize a gradient-based method to explore feature importance in deep learning feature extraction for event classification between earthquakes and explosions. Recently, \citep{clements2024grapes} showed activation feature maps to reveal the parts of the seismograms that were most contributing to predicting shaking intensity.

\begin{figure*}
\includegraphics[width=\textwidth]{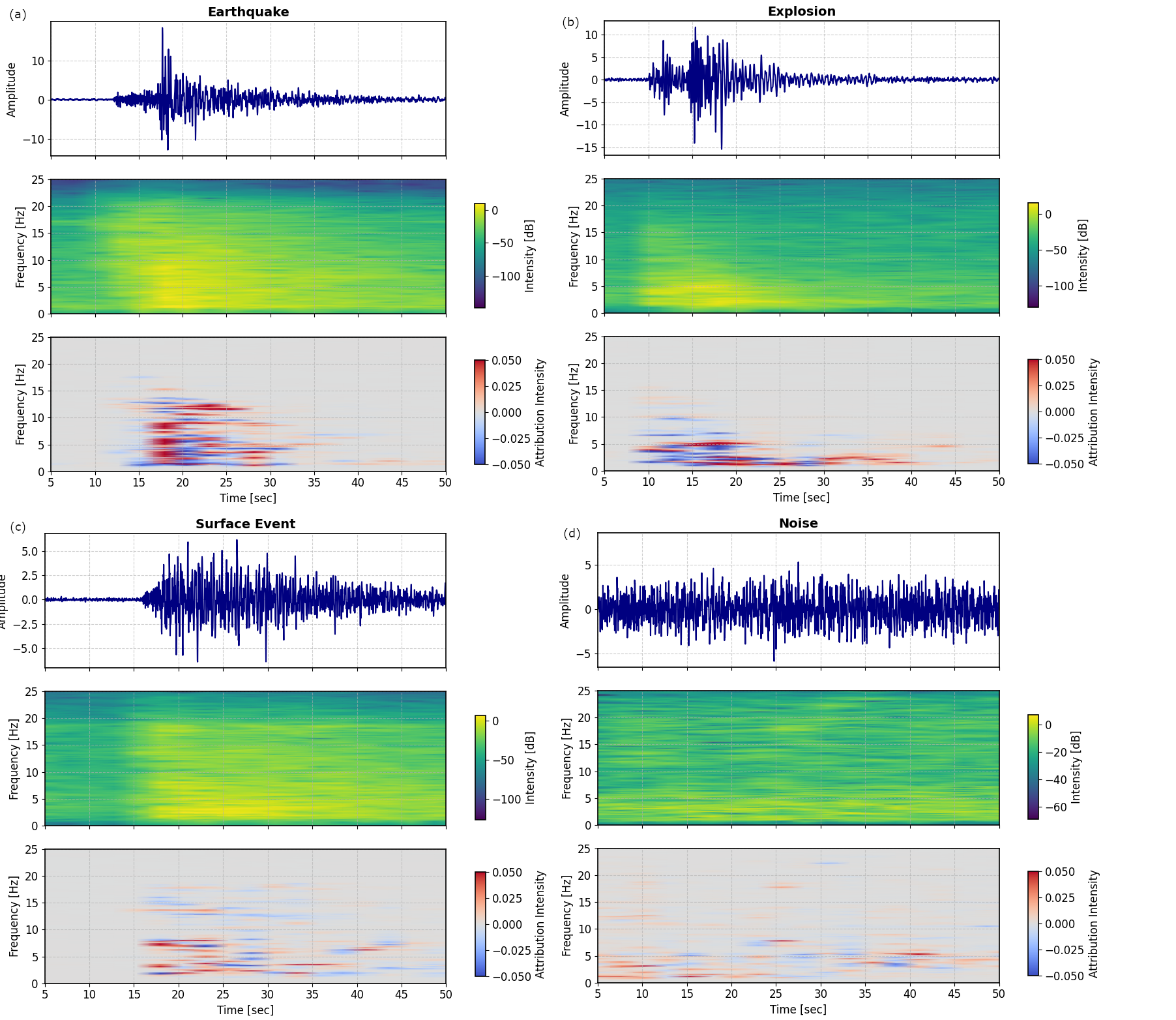}
\caption{{\bf Feature importance using the model QuakeXNet (2D) and attribution intensity.} Each subplot shows a representative waveform, spectrogram and attribution plot for each class for a) earthquakes, b) explosions, c) surface events, and c) noise.} 
\label{fig:figure9}
\end{figure*}

We chose to use the Integrated Gradients (IG) attribution method provided by the {\tt Captum} Python library, offering a robust way to interpret model predictions \citep{sundararajan2017axiomatic, alzubaidi2021review}. IG attribution assigns importance scores to input features by computing the path integral of gradients along a straight line between a baseline input and the actual input. This method ensures that the attributions satisfy key properties such as completeness and sensitivity, making it particularly suitable for complex models like neural networks \citep{sundararajan2017axiomatic, alzubaidi2021review}. We apply IG on QuakeXnet (2D) to four representative cases for each of the four classes, presenting the seismograms, spectrograms, and IG maps in Figure~\ref{fig:figure9}. For earthquakes, the most critical features are concentrated in the 5–15 Hz frequency band. They are primarily associated with the arrival of S waves, consistent with the dominance of these wave phases in earthquake signal detection and classification. In contrast, explosions exhibit significant importance in the lower frequency band of 1–5 Hz, with features predominantly synchronous with the P-wave arrivals, aligning with the known characteristics of explosions \citep[e.g.,][]{kong2022combining}, and extended duration of coda waves. Noise shows diffuse attribution across the wide range of frequencies, reflecting the broad and low-frequency nature of background or anthropogenic noise sources.

For surface events, the model highlights features primarily in the lower frequency range (below 5 Hz). Unlike earthquakes and explosions, surface events do not exhibit a clear S-wave phase, as analysts typically only pick the onset of such events \cite[e.g.,][]{ekstrom2013simple}. Instead, the identified attribution intensity may correspond to emergent or prolonged energy release patterns, which are characteristic of surface processes.  Volcanic seismicity often consists of high-frequency P and S waves from deeper sources or low-frequency (LF) events, where the frequency content is primarily controlled by the source mechanism rather than surface wave excitation \citep[e.g.,][]{chouet1996long,chouet2013multi,allstadt2014seismic,hurlimann2019debris}. It is important to note that the “surface event” classification used by the PNSN is broad, encompassing a diverse set of mass-movement and volcanic processes with correspondingly varied waveforms. This diversity likely contributes to the difficulty of training machine learning models on a single unified SU class, and should be recognized as an inherent limitation of this label.

 \section{Discussions}

This study provides a comprehensive analysis of seismic event classification through the application of CML and DL approaches, leveraging a diverse feature set derived from seismic waveforms. Our findings underscore the importance of feature selection in CML models and reveal the nuanced strengths and weaknesses of DL architectures in real-world seismic monitoring applications.

\subsection{Analysis of misclassified events}
We evaluate our best‐performing model, QuakeXNet-2D v3, on the held-out portion of the curated three-component dataset that is excluded from training (33,719 earthquakes with 109,687 traces; 3,490 explosions with 5,075 traces; 769 surface events with 902 traces; and 28,074 noise traces). For surface events, the model attains 81\% trace-level and 84\% event-level accuracy. Among the 120 misclassified surface events, 54\% (65/120) are high-confidence errors—defined as cases where a non-label class receives probability >0.9—and 18\% (22/120) remain high confidence after averaging across two stations. We forward these high-confidence candidates to PNSN analysts; the model labels most as noise, while some are identified as earthquakes. A senior PNSN analyst (>30 years of experience) confirmed that 8 of these 22 are earthquakes mislabeled as surface events in the original catalog and flagged 4 additional cases as likely shallow volcano-tectonic (VT) events. Visual inspection reveals VT-like characteristics (sharp onsets, distinct phases, short durations) despite low SNR. These findings imply that approximately 1.5–8.4\% of surface-event labels may be erroneous. For explosions, the model achieves 80\% trace-level and 84\% event-level accuracy. Of 608 misclassified events, 233 are high-confidence, including 17 flagged consistently across more than two stations; six PNSN analysts review these 17 and unanimously agree that 5 are earthquakes mislabeled as explosions, while only 4 are unanimously confirmed as true explosions, with several cases resembling volcanic deep low-frequency (DLF) events not represented in the training catalog, suggesting a potential 0.3–6\% mislabel rate for explosions. For earthquakes, the model attains 84\% trace-level and 87\% event-level accuracy. Most errors are predicted as explosions; 36\% (1,652/4,528) are high-confidence and 6\% (288/4,528) are high-confidence across multiple (>2) stations. Three senior PNSN analysts review up to 50 such cases and unanimously confirm 9 as true explosions, identify 5 as possibly DLF events from Mount Baker, and validate the remainder as earthquakes, many of which are teleseismic or proximal to known quarries. Collectively, these audits indicate that approximately 0.2–5\% of earthquake labels in the curated catalog may be incorrect.

The misclassification patterns indicate class leakage in the curated catalog rather than purely model failure. High-confidence disagreements that are often consistent across stations frequently point to surface events that behave like shallow VT earthquakes and explosions that resemble ordinary earthquakes or DLF-like signals absent from the training taxonomy. These findings imply that even modest, targeted relabeling (on the order of ~0.2–8\% by class) can influence downstream analysis of event occurence. ML classification can serve as a tool for the data curation itself by refining labels. 

% Practically, the model functions as a **catalog auditor**: we prioritize events where the model’s confidence contradicts the catalog, require multi-station agreement, batch them to analysts, and fold confirmed corrections back into the archive and retraining. The error geography (near quarries) and source-type signatures (VT/DLF candidates) motivate **hierarchical labeling**, **open-set recognition** for out-of-catalog signals, and **multi-modal fusion** with simple metadata (hour of day, depth, proximity to quarries/volcanoes). Treating high-confidence “errors” as hypotheses, not failures, yields a scalable human-in-the-loop pipeline that tightens decision boundaries and accelerates both monitoring accuracy and scientific understanding of PNW sources.

\subsection{Model Performance Relative to Other Published Models}

To the best of our knowledge, this study is the first to address classification across the three event classes most encountered at seismic networks from tectonic, anthropogenic, and geomorphologic events. Model performance in the literature is highly sensitive to choices such as the number of classes, the balance of training and testing datasets, and the architecture employed. Direct comparisons are therefore not straightforward, but placing our results in context highlights both the progress and the challenges in seismic event classification.  

Most prior work has focused on binary classification, which is inherently easier and typically yields higher performance. For example, \citet{perol2018convolutional} reported 94.9\% accuracy for discriminating earthquakes from noise, and \citet{meier2019reliable} achieved precision and recall above 99\% for the same task. Quarry blast versus earthquake discrimination has also reached near-human accuracy, with \citet{linville2019deep} obtaining accuracies above 99\% and \citet{kong2022combining} obtaining 95.2\% accuracy by combing physics-based and learned features to discriminate earthquakes and explosions. Other binary classifiers targeting noise versus earthquakes or specific volcanic signals similarly report accuracies exceeding 95–98\% \citep{wu2018seismic, chakraborty2022creime}.  

Extending to multi-class problems is more challenging. \citet{canario2020depth} achieved 96–98\% accuracy across multiple volcanic seismicity classes. More recently, \citet{maguire2024generalization} trained CNNs across diverse U.S. regions and achieved station-level accuracies of $\sim$90\% on previously unseen areas, underscoring the difficulty of robust generalization.  

Our CNN-based models achieve accuracy and F1 scores of 92–94\% on a balanced curated dataset spanning four classes, and retain high performance (F1 $\approx$0.95) under more realistic network conditions. This places our work at the high end of reported accuracies for multi-class classification, while tackling a more complex problem than most binary approaches. Importantly, our results highlight not only the feasibility of four-class discrimination but also the limits of choosing a region-specific training dataset, motivating the need for iterative dataset augmentation and generalization testing.

\subsection{Deployment on Continuous Data}

A key goal of this study is operational deployment: applying classifiers not just to curated test sets but to continuous waveform archives and real-time streams. We integrated our best deep learning model, QuakeXNet (2D) v3, into the open-source {\tt seisbench} ecosystem \citep{Woollam22}, extending its API to support multi-class classification in parallel with existing phase pickers.

Our classifier takes raw three-component seismic waveforms as input, performs preprocessing internally, and outputs four probability traces—one for each event class. The temporal resolution of these traces depends on the stride: for example, a 400-s input with a 5-s stride produces 61 probability values per class, each representing the likelihood that the following 100 s belong to that class. To convert probability traces into discrete detections, we smooth the outputs with a five-sample moving average to suppress spurious peaks. A detection is triggered when the smoothed probability exceeds 0.15,  terminated when it falls below, and validated if the within-window maximum exceeds 0.5. If multiple classes peak in the same window, the class with the highest maximum probability is assigned. We further integrated the classifier into QuakeScope \url{https://github.com/SeisSCOPED/QuakeScope}
, enabling joint operation with phase pickers on cloud-hosted data (e.g., SCEDC \citep{yu2021scedc}, NCEDC, EarthScope archives). This parallel design allows detection and discrimination to inform each other, an important feature since phase pickers are not trained on surface events. Finally, we benchmarked computational performance. QuakeXNet (2D) v3 requires only $\sim$5 s to process a full day of three-component data at 100 Hz with a 20-s stride. This cost is comparable to PhaseNet, underscoring that integrating multiple DL models into routine workflows is computationally feasible. Together, these results highlight QuakeXNet (2D) v3 as not only the most accurate but also an operationally practical solution for large-scale cataloging and real-time monitoring.

\subsection{Implications and Recommendations}

Our experiments show that while classical machine learning approaches can provide seismologically interpretable insights, they are ultimately limited compared to deep learning models. CML performance depends heavily on feature design, with Physical + Manual features emerging as the most important. Shorter windows improved earthquake recall, whereas longer windows benefited explosion and surface event detection. Broader frequency ranges consistently improved performance across classes. These findings align with seismological expectations and validate the utility of feature-based approaches for exploring signal characteristics. However, even with extensive hyperparameter tuning and feature engineering, CML models plateaued below the performance of DL classifiers and struggled to generalize to the network testing dataset.

In contrast, DL models bypass the need for intermediate feature extraction, learning directly from waveform or spectrogram representations. They consistently achieved higher accuracy, precision, and recall, while also being more computationally efficient at inference time. This efficiency makes DL approaches better suited for operational deployment, where throughput and robustness are critical.

Among the DL architectures tested, QuakeXNet 2D emerged as the most reliable model across all evaluation stages. It generalized better than SeismicCNN 2D to network and out-of-domain datasets, while maintaining high performance on curated test sets. Crucially, QuakeXNet 2D is also lightweight and fast, requiring only 70k parameters (~1.2 MB memory) and processing a full day of continuous data in ~9s. This combination of accuracy, robustness, and efficiency makes QuakeXNet 2D the recommended model for all use cases considered in this study: earthquake monitoring, explosion detection, and surface event cataloging.

Overall, these results highlight that while CML models remain valuable for interpreting seismic features, DL models—and QuakeXNet 2D in particular—provide the most practical and scalable solution for modern seismic event classification.

\begin{figure*}
\includegraphics[width=\textwidth]{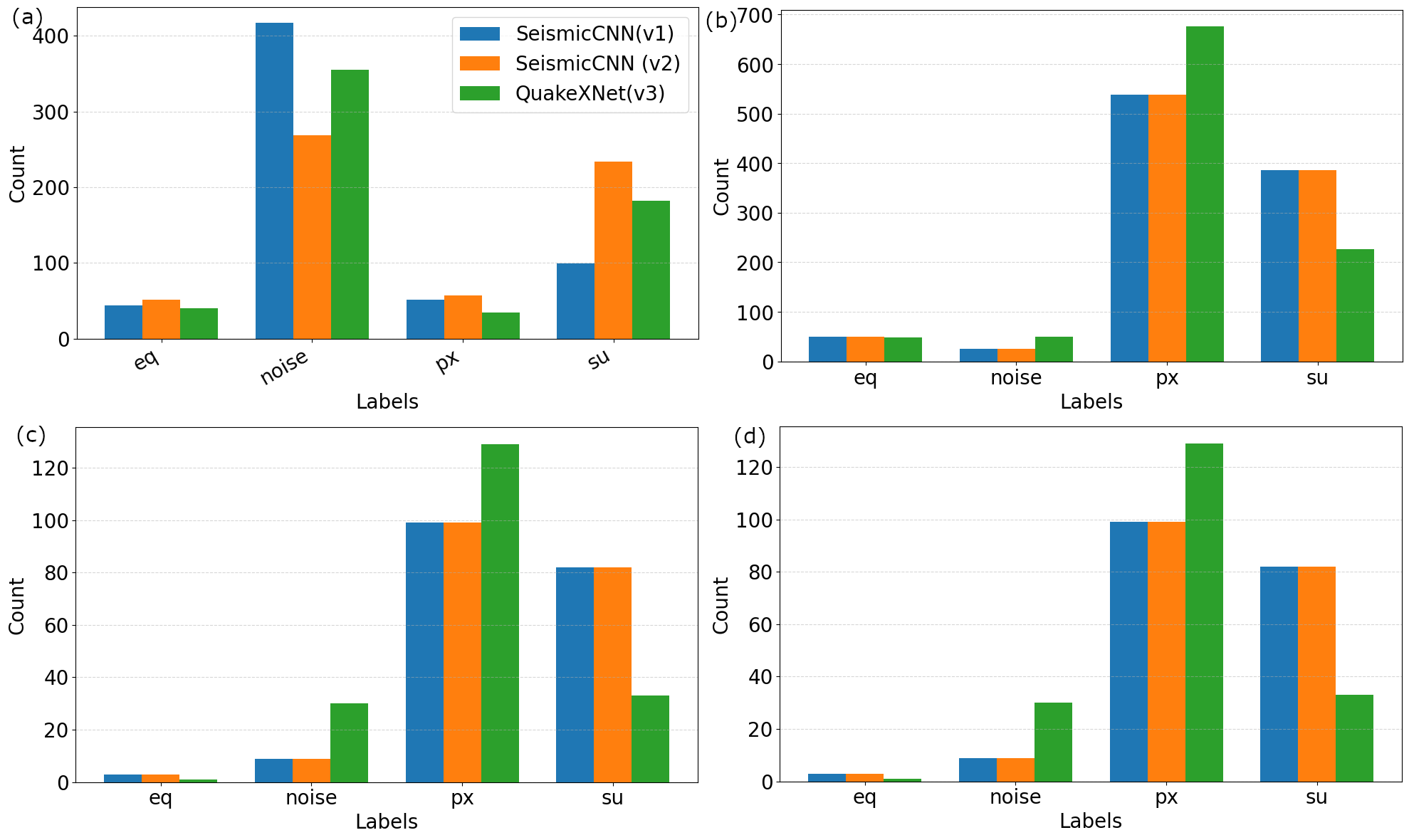}
\caption{{\bf Performance of different models on the ESEC dataset (a, c) and the near-field explosion dataset (b, d), shown at both the station level (a,b) and the event level (c,d).}} 
\label{fig:figure10}
\end{figure*}

\section{Conclusions}
In this study, we developed and evaluated classic machine learning and deep learning models for classifying dominant seismic events in the Pacific Northwest. Our results indicate that while deep learning models, such as QuakeXNet (2D) and SeismicCNN (2D), perform well on test datasets and offer higher classification confidence for the Pacific Northwest data, classic machine learning models have also demonstrated relatively good performance and can be used in an ensemble fashion.

Our analysis highlighted the key features that distinguish different seismic event types. Classic machine learning models emphasized kurtosis-based features, which provided clear separation among noise, surface events, explosions, and earthquakes. In particular, kurtosis in specific frequency bands (e.g., 3–10 Hz) proved highly informative, while contextual features such as time of day helped discriminate explosions from other classes. Deep learning attribution maps offered complementary insights, showing that earthquakes are characterized by energy concentrated in the 5–15 Hz band and linked to S-wave arrivals, whereas explosions emphasize lower-frequency energy (1–5 Hz) associated with P-wave onsets and extended codas. Noise showed broad, diffuse patterns, while surface events exhibited energy concentrated at low frequencies without distinct S-wave phases, reflecting emergent or prolonged release processes. Together, these findings demonstrate that both engineered and learned features converge on physically meaningful signal properties, and that integrating multiple perspectives improves our ability to discriminate between event types.

Our analysis also shows that combining data from multiple stations improves classification performance by averaging out noise and reducing the impact of individual station biases, another common form of network seismology and association. Deep learning models performed better than Classic Machine Learning models both in terms of classification performance as well as computational costs on the curated PNW dataset. Further, we enhanced the generalizability of the original deep learning models by training them with an incrementally enriched out-of-domain dataset. Overall, we found that QuakeXNet(2D) - v3 is the best model in terms of performance, computational costs, and generalizability. This model performed well on the curated PNW datasets and out-of-domain surface events and explosions, and should be utilized for large-scale detection of surface events.

\begin{acknowledgements}
We acknowledge support from the US Geological Survey Earthquake Science Center through  Cooperative Agreement G23AC00278 and the partial funding provided by the PNSN (USGS cooperative agreement G20AC00035). We thank seismic analyst volunteers Amy Wright, Paul Bodin, and Barrett Johnson for their help validating event categories.

\end{acknowledgements}

\section{Data and code availability}
The seismic waveform dataset used in this study is publicly accessible at \url{https://github.com/EarthML4PNW/PNW-ML }\citep[e.g.,][]{ni2023curated}.
All models evaluated in this study, including feature extraction scripts and hyperparameter tuning configurations, are available in the repository \url{https://github.com/Akashkharita/PNW\_Seismic\_Event\_Classification}. Real-time testing and deployment codes can be found at \url{https://github.com/Akashkharita/Surface\_Event\_Detection}.
Additionally, trained models are archived in the Zenodo repository at \url{https://zenodo.org/records/13334838}, allowing users to reproduce the workflow or apply the models to other regions of interest.

\section*{Competing interests}
The authors declare no competing interests.

%% If the article is accepted, a separate bibfile must be uploaded along with the compiled manuscript, source file, and separate figure files.
%% When available, DOI numbers must be provided for all references, including datasets and codes. 
\bibliography{mybibfile}

\end{document}

\maketitle

\section{Scatnet Feature Extraction}
The scattering coefficients are derived using a two-layer scattering network designed to capture hierarchical, multi-scale representations of the seismic waveforms. The network relies on Morlet wavelets as its basis functions, decomposing the input signal into its time-frequency components. The parameters of the scattering network are meticulously chosen to balance resolution and sensitivity to signal variations.

In the first layer, the octaves parameter determines the number of dyadic frequency bands analyzed. For instance, with five octaves, the frequency range is divided into progressively narrower bands, enabling the network to detect both low- and high-frequency components. The resolution parameter, set to two, dictates the granularity of the frequency band division within each octave, increasing the frequency resolution. The quality factor, set to one in the first layer, controls the trade-off between time and frequency localization, ensuring sharp temporal responses for precise event detection.

The second layer builds upon the first by considering interactions between frequency bands captured in the first layer. Here, the octaves, resolution, and quality are redefined (e.g., quality set to three), allowing the layer to focus on broader and smoother modulations of the signal. This layer computes coefficients that encode second-order interactions, representing how different frequency bands influence each other over time. These parameters are particularly crucial for capturing non-linearities and complex seismic event signatures, such as those arising from mixed-source phenomena or overlapping events.

The scattering coefficients at each layer are computed as the modulus of the wavelet transform, aggregated over time using operations like max pooling to summarize the signal's behavior over each time segment. The total number of wavelets (e.g., octaves × resolution) and their specific bandwidths ensure that the scattering coefficients offer a rich representation of the signal's energy distribution and dynamics across scales. These coefficients are normalized using logarithmic scaling, ensuring stability and robustness, particularly for signals with wide-ranging amplitudes. This parameterization allows the scattering network to effectively capture the complex, multi-scale features essential for characterizing seismic events in diverse geohazard contexts.

\section{Hyper-parameter tuning of CML models}

In this study, we evaluated the performance of various machine learning algorithms on a dataset comprising 1000 traces per class, utilizing physical and TSFEL-generated features. For each algorithm, we conducted an exhaustive search over all possible hyperparameter combinations defined within a grid for each respective model. We employed five-fold cross-validation for every hyperparameter combination to ensure robustness, recording the resulting F1 scores to assess model performance.

\paragraph*{MLP (Multi-Layer Perceptron)}
To tune its performance, the Multi-Layer Perceptron (MLP) was configured with several hyperparameters. The \texttt{hidden\_layer\_sizes} parameter specifies the number of neurons in each hidden layer, such as \texttt{(100)}, indicating one hidden layer with 100 neurons. Larger layer sizes may enhance the model's capacity to capture complex patterns, but can lead to overfitting if not tuned properly. The \texttt{activation} function was set to either \texttt{relu} (Rectified Linear Unit) or \texttt{tanh} (Hyperbolic Tangent). While \texttt{relu} offers efficient learning by setting negative inputs to zero, \texttt{tanh} maps inputs to a range between -1 and 1, which can help when centered outputs are needed. The \texttt{solver} for optimization was fixed to \texttt{adam}, a widely used method that combines the advantages of RMSProp and Stochastic Gradient Descent (SGD). Finally, the \texttt{max\_iter} parameter was set to \texttt{500}, limiting the maximum number of iterations for convergence during training.

Grid for MLP

\begin{verbatim}
    
mlp_param_grid = {`hidden_layer_sizes': [(100,), (200,), (300,)],
 `activation': [`relu', `tanh'],
 `solver': [`adam'],
 `max_iter': [500]}

\end{verbatim}

Optimal hyperparameter values for MLP were found to be as follows -

\begin{verbatim}
{`activation': `relu', `hidden_layer_sizes': (200,), `max_iter': 500, 'solver': `adam'}
\end{verbatim}

\paragraph*{SVC (Support Vector Classifier)}
The Support Vector Classifier (SVC) employed two key hyperparameters for tuning. The \texttt{C} parameter controls the regularization strength, with smaller values enforcing stronger regularization to reduce overfitting and larger values allowing for more flexible decision boundaries. The \texttt{kernel} parameter determines how the algorithm maps the input features into higher-dimensional spaces. Two kernel options were explored: \texttt{linear}, which fits a hyperplane to separate classes in the original feature space, and \texttt{rbf} (Radial Basis Function), which uses a Gaussian-based method for projecting data into higher dimensions, enabling the classifier to handle non-linear separable data.

Grid for SVC
\begin{verbatim}
svc_param_grid = {`C': [0.1, 1, 10], `kernel': [`linear', `rbf']}
\end{verbatim}

Optimal hyperparameter values for SVC were found to be as follows - 
\begin{verbatim}
{`C': 10, `kernel': `rbf'}
\end{verbatim}

\paragraph{KNN (K-Nearest Neighbors)}
The K-Nearest Neighbors (KNN) algorithm was tuned using the \texttt{n\_neighbors} parameter, which specifies the number of neighbors to consider when making classification decisions. Smaller values result in more localized decision boundaries, potentially increasing sensitivity to noise, while larger values produce smoother decision surfaces.

Grid for KNN
\begin{verbatim}
knn_param_grid = {`n_neighbors': [3, 5, 7, 9]} 
\end{verbatim}

Optimal hyperparameter values for KNN were found to be as follows - 
\begin{verbatim}
{`n_neighbors': 7}
\end{verbatim}

\paragraph{Logistic Regression (LR)}
Logistic Regression was configured with three hyperparameters. The \texttt{C} parameter, representing the inverse of regularization strength, was explored over a range of values. Smaller values enforce stronger regularization to mitigate overfitting, whereas larger values allow the model to capture more intricate relationships. The \texttt{penalty} was fixed to \texttt{l2} (Ridge regularization), which adds a squared magnitude of coefficients as a penalty term. Two solvers were considered for optimization: \texttt{lbfgs}, a quasi-Newton method efficient for small- to medium-sized datasets, and \texttt{liblinear}, which works well for L2-regularized problems in smaller datasets.
Grid for LR

\begin{verbatim}
lr_param_grid = {`C': [0.01, 0.1, 1, 10, 100], `penalty': [`l2'], `solver': [`lbfgs', `liblinear']}
\end{verbatim}

Optimal hyperparameter values for LR were found to be as follows - 
 \begin{verbatim}
 {'C': 0.01, 'penalty': 'l2', 'solver': 'lbfgs'}
\end{verbatim}

\paragraph{Random Forest (RF)}

The Random Forest model was tuned using two main parameters. The \texttt{n\_estimators} parameter determines the number of decision trees in the forest. Increasing this number typically improves accuracy but also adds to the training time. The \texttt{max\_depth} parameter defines the maximum depth of each tree. Deeper trees can better model complex relationships but may overfit the training data.

Grid for RF
\begin{verbatim}
rf_param_grid = {`n_estimators': [100, 200, 300], `max_depth': [None, 10, 20]}
\end{verbatim}

Optimal hyperparameter values for RF were found to be as follows - 
\begin{verbatim}
{`max_depth': None, `n_estimators': 300}
\end{verbatim}

\paragraph{XGBoost (XGB)}

For the XGBoost algorithm, the hyperparameters \texttt{n\_estimators} and \texttt{max\_depth} were optimized. The \texttt{n\_estimators} parameter specifies the number of boosting rounds or trees. Higher values generally improve performance but increase computational cost. The \texttt{max\_depth} parameter limits the depth of individual trees, with deeper trees capable of capturing more intricate patterns at the expense of a higher risk of overfitting.

Grid for XGB

\begin{verbatim}
xgb_param_grid ={`n_estimators': [50, 100, 150], `max_depth': [3, 5, 7]}
\end{verbatim}

Optimal hyperparameter values for XGB were found to be as follows - 
\begin{verbatim}
{`max_depth': 7, `n_estimators': 100}
\end{verbatim}

\paragraph{LightGBM (LGBM)}

The LightGBM model was configured with three hyperparameters. The \texttt{n\_estimators} parameter, similar to XGBoost, determines the number of boosting rounds or trees. The \texttt{max\_depth} parameter restricts the depth of the trees, with higher values enabling the model to learn more complex patterns. Finally, the \texttt{num\_leaves} parameter sets the maximum number of leaves per tree, with a default of 31, balancing model complexity and computational efficiency.

By systematically exploring these hyperparameter grids, the study aimed to determine the optimal configuration for each model, maximizing F1 scores and enhancing classification performance on the multi-class dataset.

Grid for LGBM
\begin{verbatim}
lgbm\_param\_grid = {`n_estimators': [10, 50, 100], `max_depth': [3, 5], `num_leaves':[7]}
\end{verbatim}

Optimal values LGBM were found to be as follows - 
\begin{verbatim}
{`max_depth': 3, `n_estimators': 50, `num_leaves': 7}
\end{verbatim}

\begin{figure*}
\includegraphics[width=\textwidth]{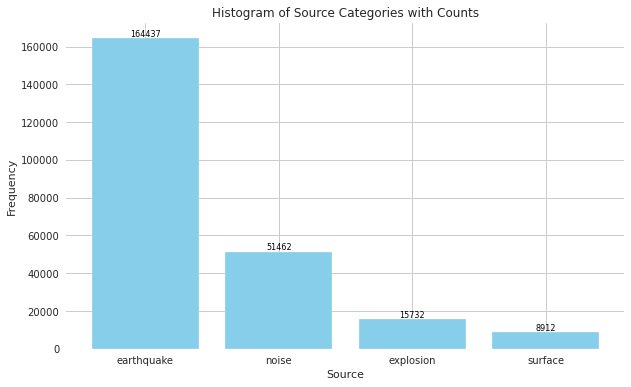}
\caption{Number of waveforms available from \citet{ni2023curated}.}
\label{fig:figure_supp1}
\end{figure*}

\begin{figure*}
\includegraphics[width=\textwidth]{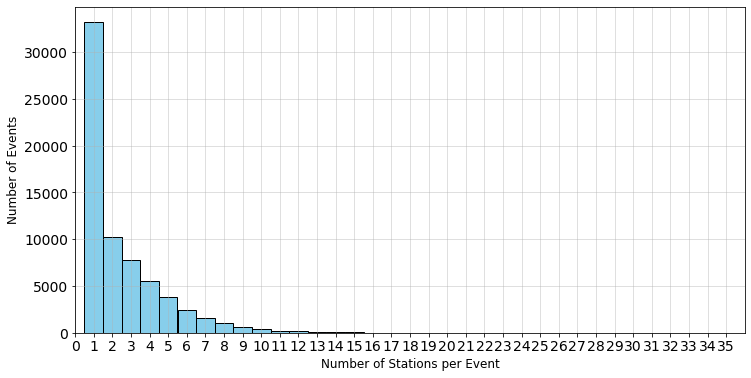}
\caption{{\bf Number of stations per event} Most of the events were detected at one or two stations.}
\label{fig:figure_supp2}
\end{figure*}

\begin{figure*}
\includegraphics[width=\textwidth]{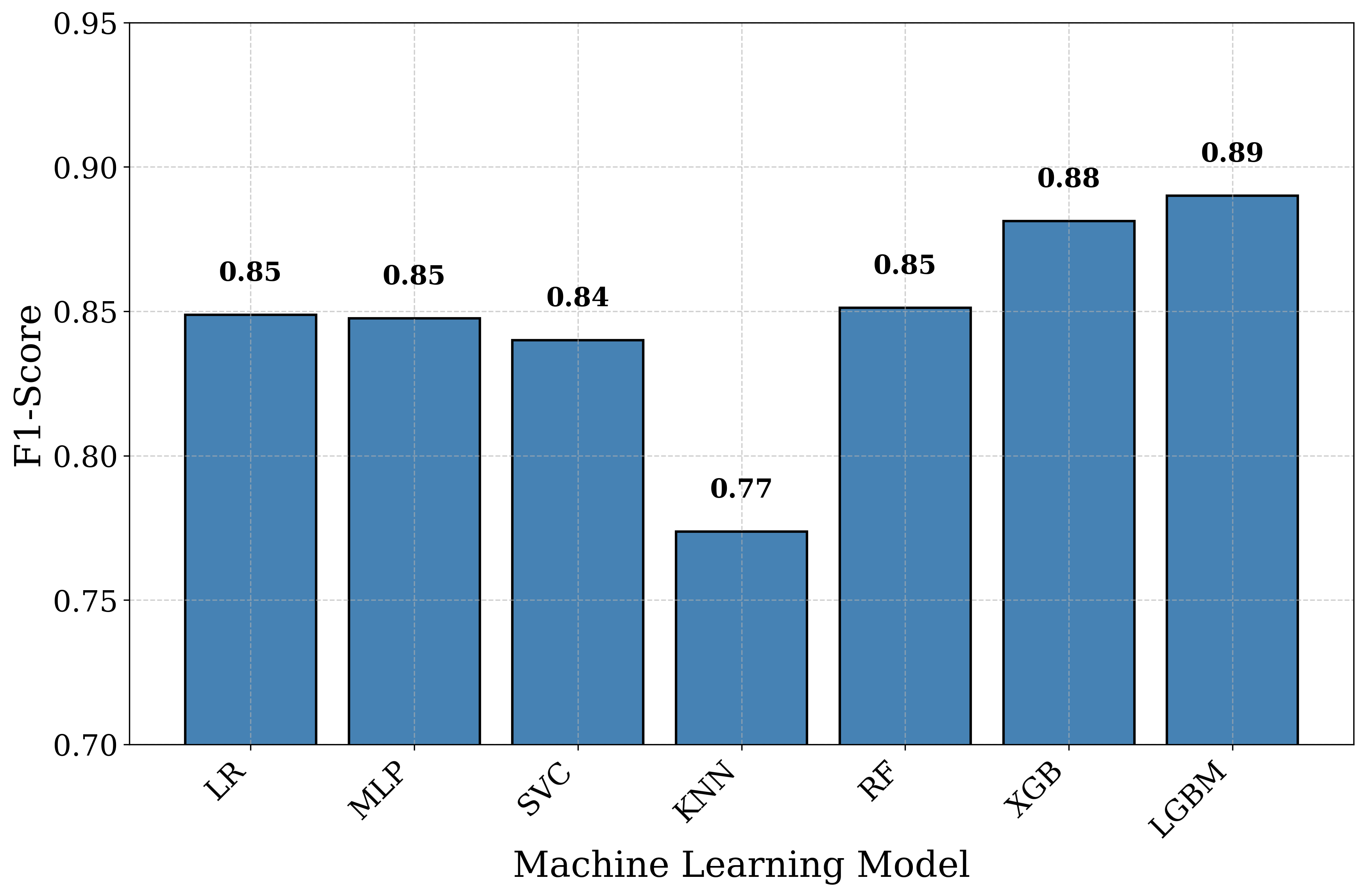}
\caption{{\bf F1 Score} of CML models tested and estimated on the test data from the curated data set.}
\label{fig:figure_supp3}
\end{figure*}

\begin{figure*}
\includegraphics[width=\textwidth]{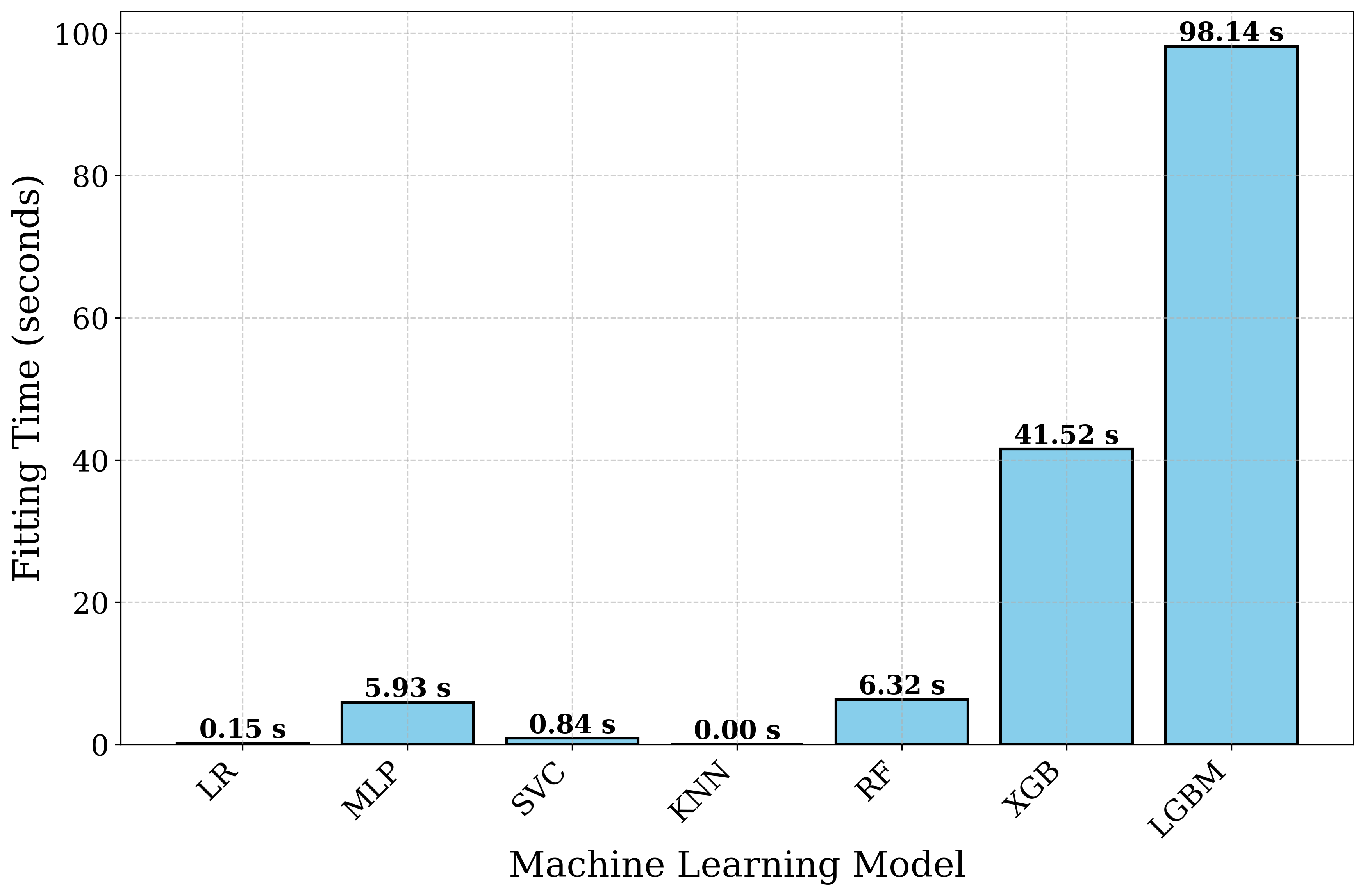}
\caption{{\bf Computational time for training the CML models.}}
\label{fig:figure_supp4}
\end{figure*}

\begin{figure*}
\includegraphics[width=\textwidth]{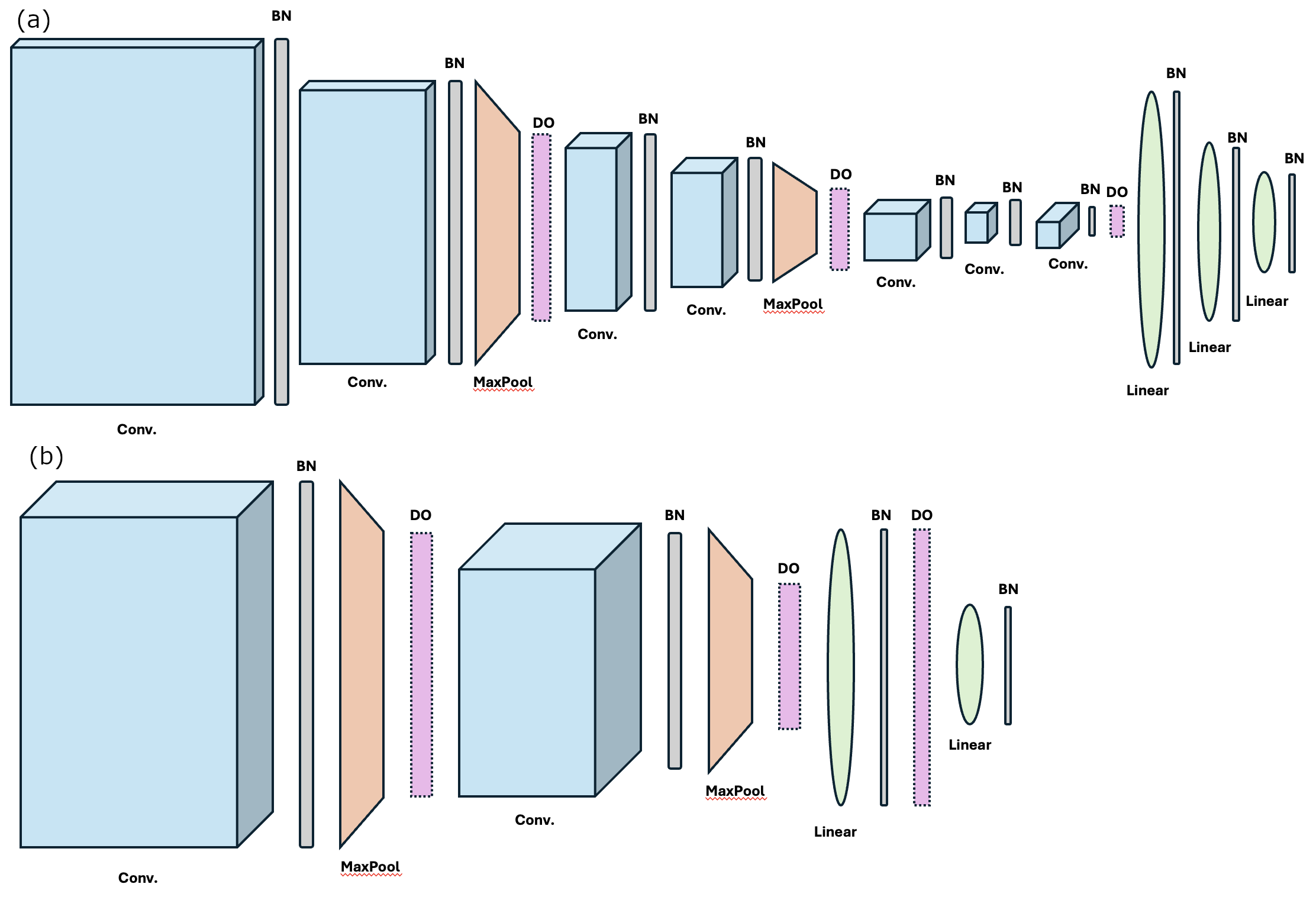}
\caption{{\bf Neural network architectures} depicting two types of canonical forms used in this study (a) QuakeXNet and (b) SeismicCNN.}
\label{fig:figure_supp5}
\end{figure*}

\begin{figure*}
\includegraphics[width=\textwidth]{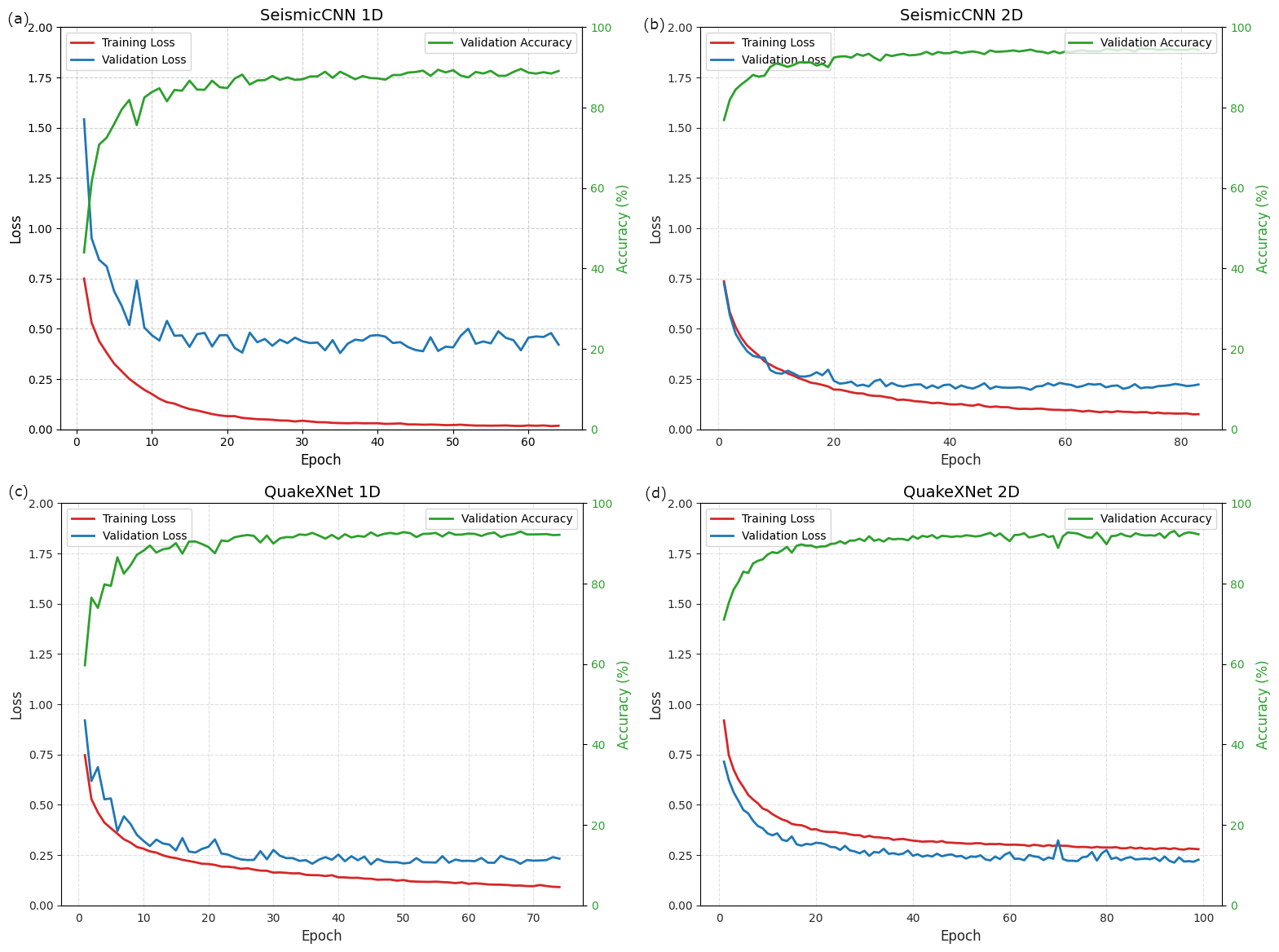}
\caption{{\bf Training and validation accuracy and loss curves} for (a) QuakeXNet (1D), (b) QuakeXNet (2D), (c) SeismicCNN (1D) and (d) SeismicCNN (2D).}
\label{fig:figure_supp6}
\end{figure*}

\begin{figure*}
\includegraphics[width=\textwidth]{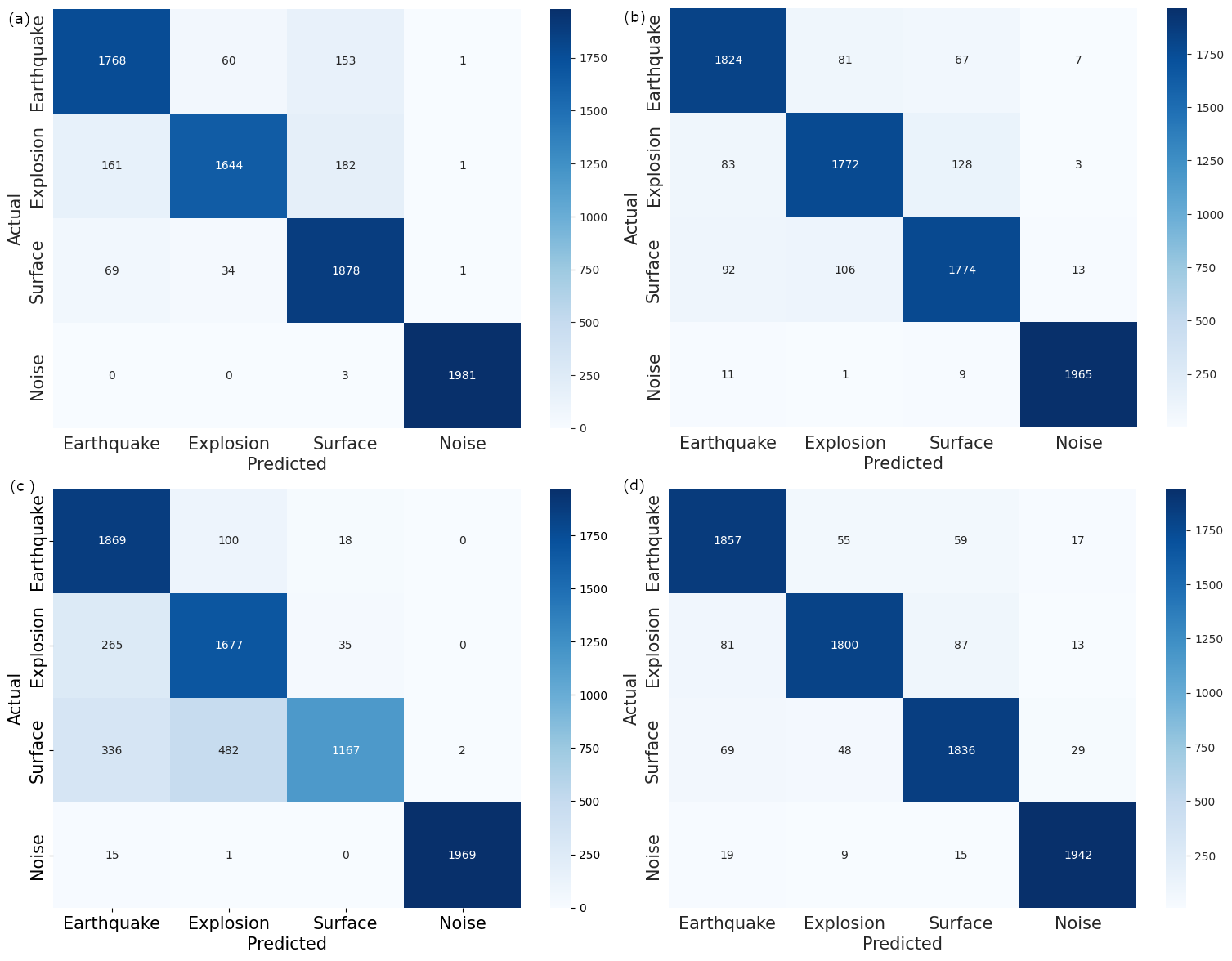}
\caption{{\bf Confusion Matrix of the four DL models} on the validation set from the curated data sets with (a) QuakeXNet (1D), (b) QuakeXNet (2D), (c) SeismicCNN (1D), (d) SeismicCNN (2D). These confusion matrices highlight where the confusion is among classes.} 

\label{fig:figure7}
\end{figure*}

\begin{figure*}
\includegraphics[width=\textwidth]{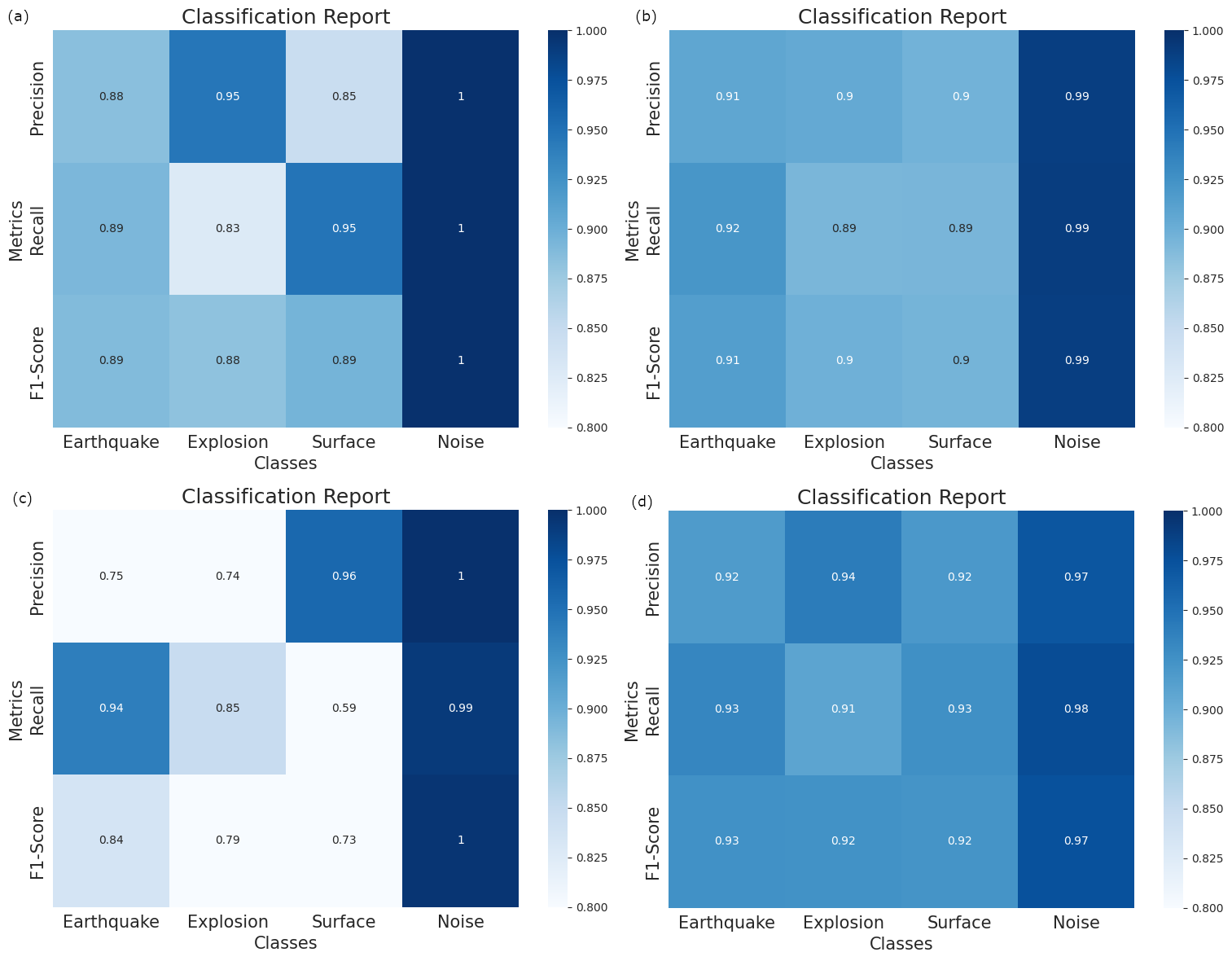}
\caption{{\bf Classification report of the four DL models} on the validation set from the curated data sets with (a) QuakeXNet (1D), (b) QuakeXNet (2D), (c) SeismicCNN (1D), (d) SeismicCNN (2D).} 

\label{fig:figure8}
\end{figure*}

\begin{figure*}
\includegraphics[width=\textwidth]{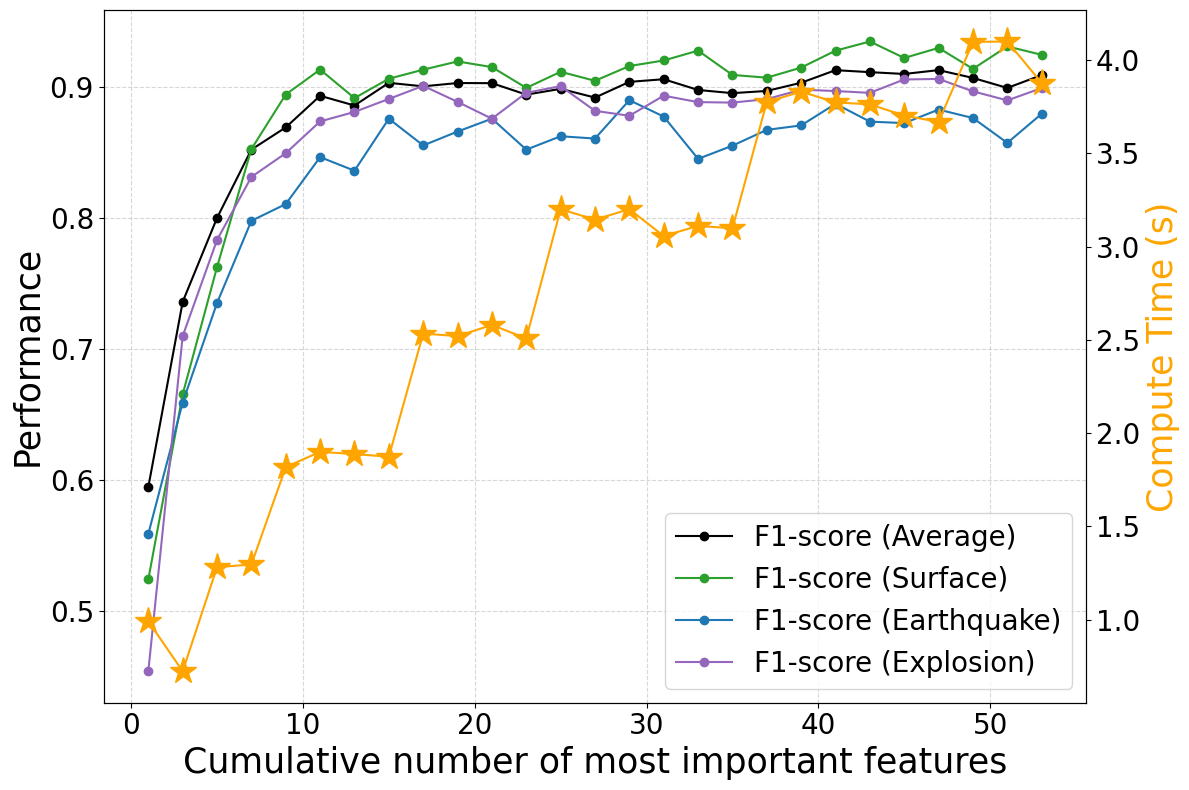}
\caption{{\bf Performance variation of each class} with cumulatively increasing number of most important features, training time on the twin axis.} 

\label{fig:figure9}
\end{figure*}

\begin{figure*}
\includegraphics[width=\textwidth]{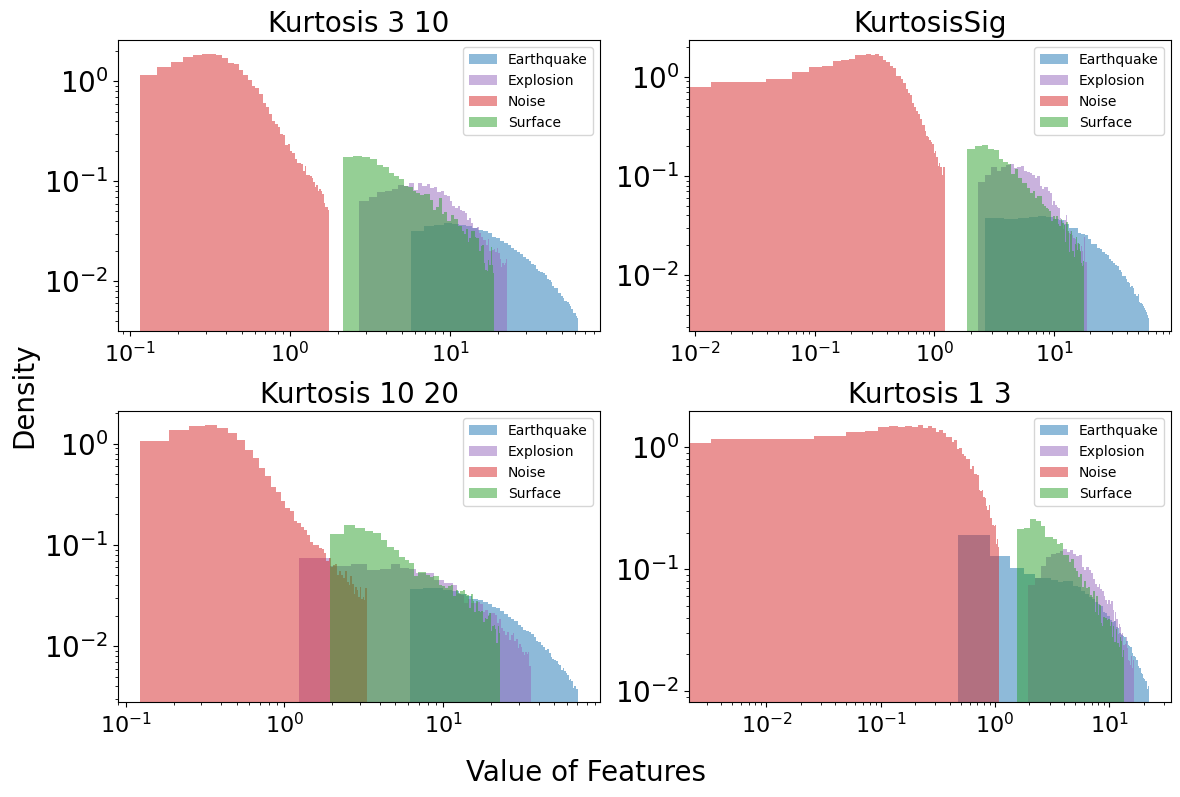}
\caption{{\bf Histograms of distribution of the most important features} for each class.} 

\label{fig:figure10}
\end{figure*}

\begin{figure*}
\includegraphics[width=\textwidth]{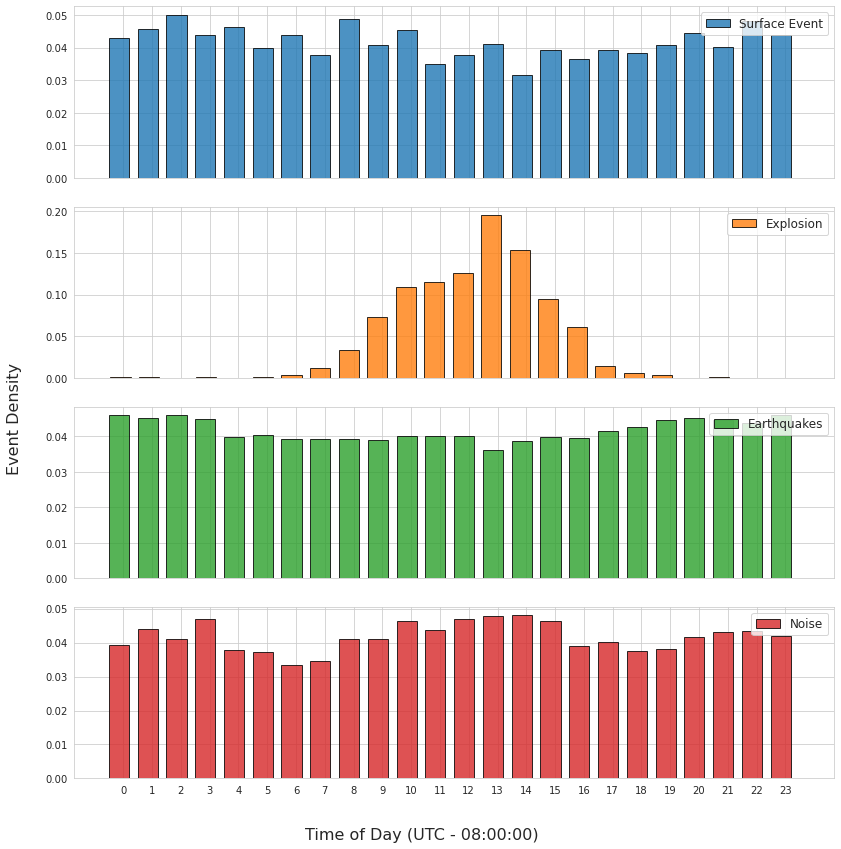}
\caption{{\bf Hour of the day distributions for all the classes}, this feature was very important for classification.} 

\label{fig:figure11}
\end{figure*}

\begin{figure*}
\includegraphics[width=\textwidth]{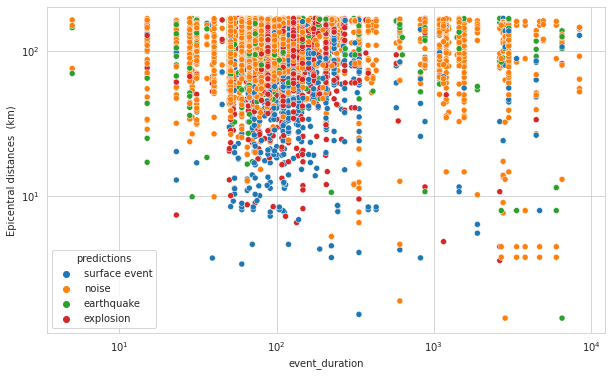}
\caption{{\bf Event classification for IRIS ESEC events} with respect to epicentral distance and event duration.} 

\label{fig:figure12}
\end{figure*}

\begin{figure*}
\includegraphics[width=\textwidth]{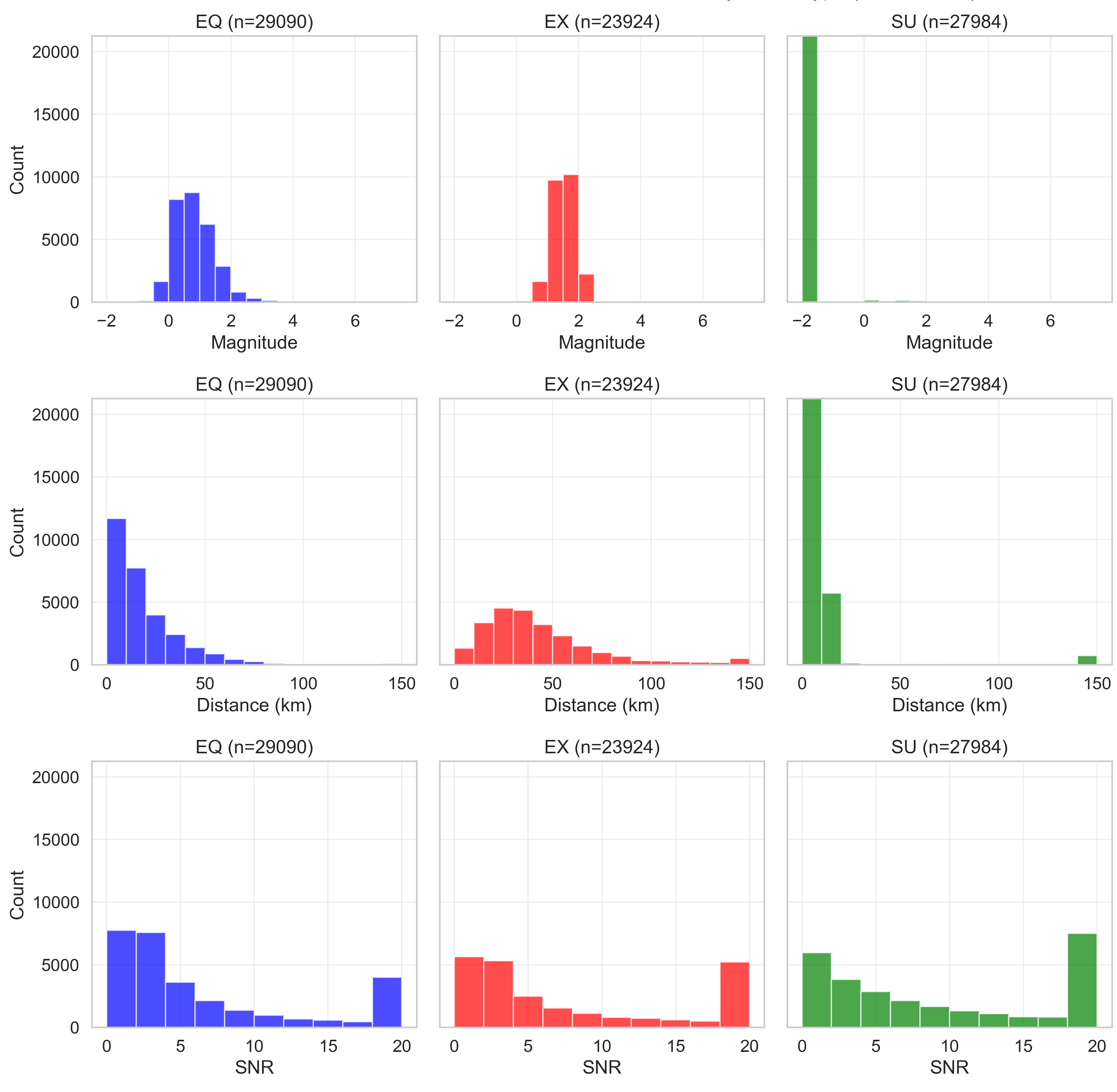}
\caption{{\bf Distribution of Network Testing data}: with distribution of the magnitudes (top panel), source-receiver range (middle panel), signal-to-noise ratio (bottom panels) and earthquakes (left column and in blue), explosions (middle column and in red), and surface events (in green). } 
\label{fig:figure13}
\end{figure*}

\begin{figure*}
\includegraphics[width=0.7\textwidth]{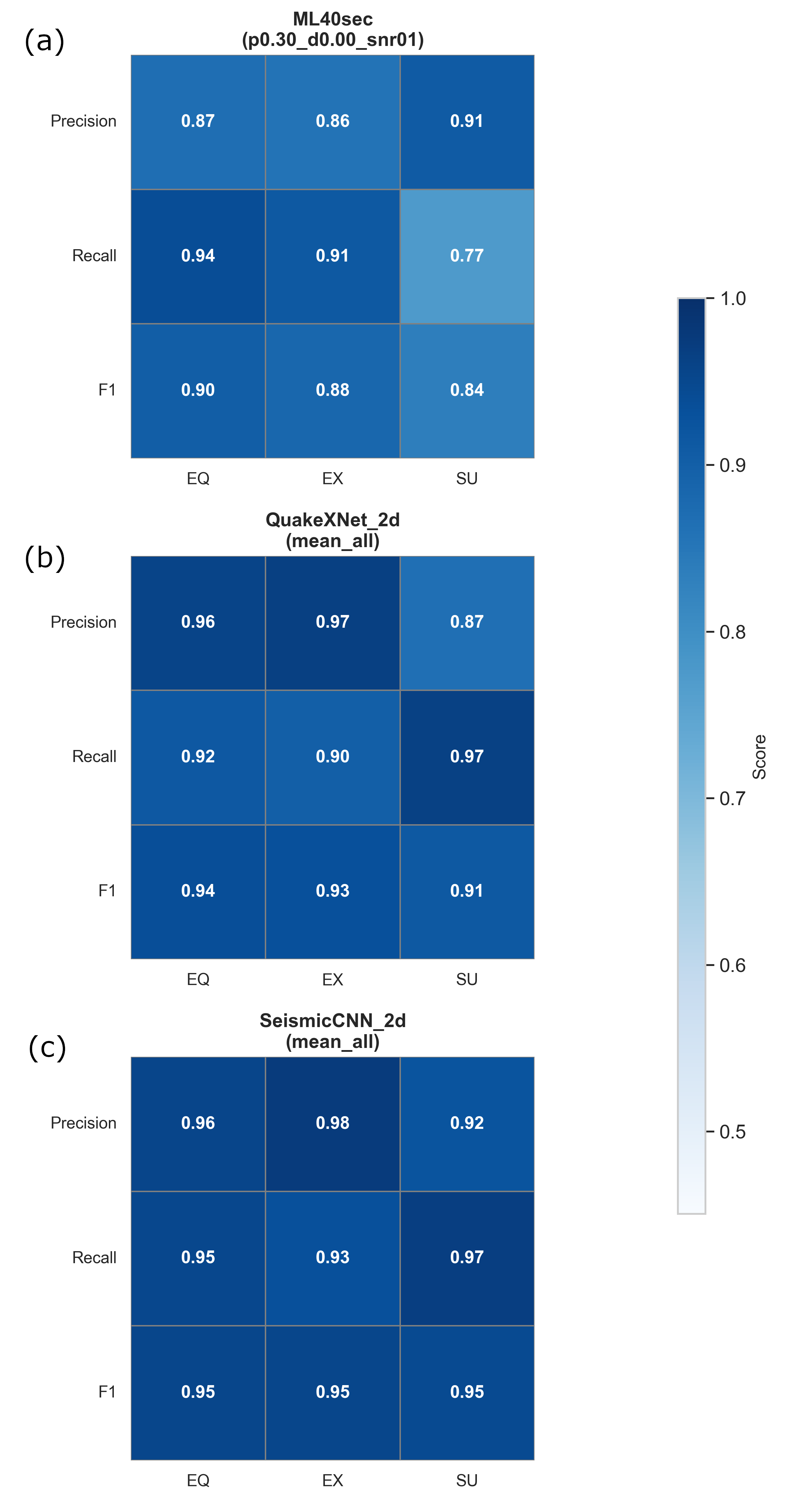}
\caption{{\bf Classificaiton report on the network testing data } for three best models: a) ML40sec, b) QuakeXNet\_2D, and c) SeismicCNN\_2D. These are network-average scores and demonstrate the relative performance among classes on the network test data.} 
\label{fig:figure14}
\end{figure*}

\begin{figure*}
\includegraphics[width=\textwidth]{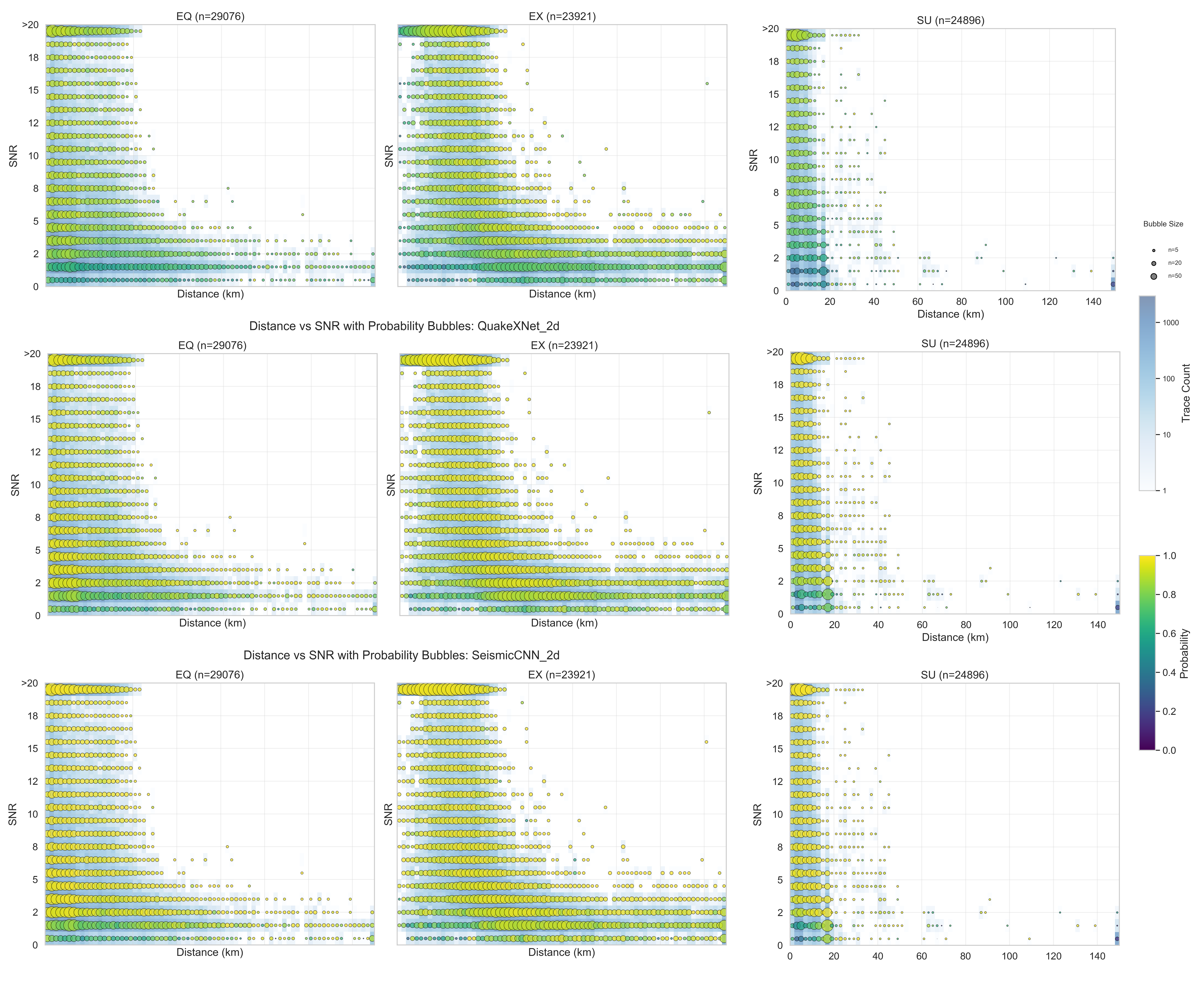}
\caption{{\bf Impacts of SNR and Distance over performance}. We show} 
\label{fig:figure15}
\end{figure*}

\bibliography{mybibfile}